\documentclass[12pt]{amsart}

\usepackage{amssymb,amsmath,graphicx,color,textcomp, amsthm,bbm,bbold, enumerate,booktabs}
\usepackage{chngcntr}
\usepackage{apptools}
\usepackage{mathrsfs}%
\usepackage{float}
\AtAppendix{\counterwithin{theorem}{section}}
\definecolor{Red}{cmyk}{0,1,1,0}

\definecolor{verde}{cmyk}{1,0,1,0}

\definecolor{loka}{cmyk}{.5,0,1,.5}
\definecolor{azul}{cmyk}{1,1,0,0}

\newcommand{\cplex}{{\rm I\hspace{-6pt}C}}
\usepackage{subfigure}%

\numberwithin{equation}{section}

\newtheorem{teo}{Theorem}[section]
\newtheorem{defi}[teo]{Definition}
\newcommand{\real}{{\rm I\!R}}
\newcommand{\nat}{{\rm I\!N}}
\renewcommand{\sin}{\hspace{2pt}\textrm{sin}}
\renewcommand{\cos}{\hspace{2pt}\textrm{cos}}

\newcommand{\be}{\begin{equation}}
\newcommand{\ee}{\end{equation}}

\begin{document}
\title{Complete Monotonicity of Fractional Kinetic Functions}
\author{Ester C. F. A. Rosa$^1$}
\address{$^1$ Department of Applied Mathematics, Institute of Mathematics,
 Statistics and Scientific Computation, University of Campinas --
UNICAMP, rua S\'ergio Buarque de Holanda 651,
13083--859, Campinas SP, Brazil\newline
e-mail: {\itshape \texttt{ra991706@ime.unicamp.br, capelas@ime.unicamp.br }}}

\author{E. Capelas de Oliveira$^1$}

\begin{abstract}The introduction of a fractional differential operator defined in
	terms of the Riemann-Liouville derivative makes it possible to
	generalize the kinetic equations used to model relaxation in
	dielectrics. In this context such fractional equations are called
	fractional kinetic relaxation equations and their solutions, called
	fractional kinetic relaxation functions, are given in terms of
	Mittag-Leffler functions. These fractional kinetic relaxation
	functions generalize the kinetic relaxation functions associated with
	the Debye, Cole-Cole, Cole-Davidson and Havriliak-Negami models, as the
	latter functions become particular cases of the fractional solutions,
	obtained for specific values of the parameter specifying the order of
	the derivative.  
\\
The aim of this work is to analyse the behavior of these fractional functions
in the time variable. As theoretical tools we use the theorem by Bernstein on
the complete monotonicity of functions together with Titchmarsh's inversion
formula.  The last part of the paper contains the graphics of some of those
functions, obtained by varying the value of the parameter in the fractional
differential operator and in the corresponding Mittag-Leffler functions. The
graphics were made with Mathematica 10.4.  
\newline 
\end{abstract}
\maketitle
\section{Introduction}

The complete monotonicity of functions has been the subject of studies carried
out between 1920 and 1930 by S. Bernstein, F. Hausdorff and V. Widder, a
problem then known as moment problem
\cite{Bernstein:1929,Hausdorff:1921,Widder:1931}. Completely monotonic
functions are non-negative functions which have derivatives of all orders with alternating signs. Reference \cite{Widder:1946} presents a detailed report on the properties of completely
monotonic functions and their characterization while Feller \cite{Feller:1970}
discusses the complete monotonicity of functions by means of their relation
with infinitely divisible measures.  Completely monotonic functions have
remarkable applications in several branchs of science. For instance, they play
important roles in potential theory, probability theory, physics, combinatorics
and in numerical and asymptotic analysis \cite{Li:2013,Lokenath:2003,
Machado:2010}. 

In recent times, several authors have shown that many functions defined in
terms of gamma, poligamma and other special functions are completely monotonic
and used this fact to deduce new interesting inequalities
\cite{Alzer:1997,Berg:2001,Ismail:1986,Merkle:2014}. We can find in the literature some texts in which one discusses the complete monotonicity
of Mittag-Leffler functions \cite{Miller:1997,Samko:2001,An:2012,Vaz:2014,Garrapa:2015,Simon:2015}, a class of functions emerging in fractional calculus
\cite{Machado:2011}.

We begin this work considering a Mittag-Leffler function with two parameters.
We then choose adequate  ranges of values for those parameters and discuss
the monotonicity of the resulting function. 

Next, we use Mori-Zwanzig's rigorous theory  of projection operators to present
the equation governing the temporal evolution of the correlation function
\cite{Popov:2014}.  This equation is one of the main results of the many-body
problem of statistical mechanics, besides playing a fundamental role in the
analysis of the dynamic behavior of fluids. 

We solve, in this work, the kinetic relaxation equations and find
their solutions, which are relaxation functions in the time variable and which are given
in terms of Mittag-Leffler functions. This is really an advancement since the available
models were described only in the frequency domain. 

Nowadays, there is a natural tendency in mathematics to generalize differential equations,
whether ordinary or partial. This kind of generalization takes place in a variety of ways,
according to which of the various fractional derivatives defined in literature is employed
\cite{Podlubny:1999,Hilfer:2000,Capelas:2015,Machado:2014}. 

Mainardi and Gorenflo generalize Newton's classical model of linear
viscoelasticity replacing its first order derivative with a fractional
derivative of order $\upsilon \in (0,1)$ \cite{Mainardi:2007}. In the same
paper they also discuss the generalization of the (exponential) relaxation
model. In discussing these two fractional models, both fractional derivatives,
Caputo and Riemann-Liouville, are compared, as well as the initial conditions
of each model. 

Relaxation processes in dielectrics have also been discussed with the help of
fractional calculus \cite{Uchaikin:2003,Novikov:2005,Garrappa:2016}. In \cite{Baleanu:2011},
the authors apply an inovative numerical algorithm to find the solution of the
fractional equation for the response function associated with the model of
Havriliak-Negami.

In \cite{Ester:2015} we defined fractional differential equations which
generalize kinetic relaxation equations by introducing in them derivatives of
fractional order, i.e.  by replacing the integer parameter in their derivatives
with a parameter assuming values in $(0,1]$ and using the Riemann-Liouville
definition of fractional derivative in place of classical, integer order
derivatives. The resulting equations are called fractional kinetic relaxation
equations; their solutions will be given in terms of Mittag-Leffler functions.
We are thus naturally led to investigate the behavior of such solutions with
respect to time and that is what we do here. 

In section two we define a completely monotonic function and present
Bernstein's theorem and Titchmarsh's formula, together with some examples of
completely monotonic functions. In section three we discuss the complete
monotonicity of the three-parameter Mittag-Leffler function when one of its
parameters is equal to unit. This study begins with the graphical analysis of
the behavior of its distribution function as one varies its other two, free
parameters. A more rigorous analysis is then carried using a theorem that
relates the complete monotonicity of functions to the properties of their
Laplace transforms \cite{Gripenberg:1990}.  In section four we solve the
kinetic relaxation equations and present the graphics of their solutions.  We
then study, in section five, the complete monotonicity of these kinetic
relaxation functions.  Section six is about the complete monotonicity of the
fractional kinetic relaxation functions, solutions of fractional kinetic
relaxation equations. We present our conclusions in section seven.

\section{Preliminaries}

In order to be able to discuss the complete monotonicity of solutions
of kinetic relaxation equations and of their fractional versions, we
present in this section the definitions, theorems and formulas we shall use
to conduct this analysis. A particularly important result is the Bernstein
theorem, which relates function monotonicity to the sign of the spectral
distribution function associated with the function and thus reduces the
problem to an inequality.

\begin{defi} We say that a function $f(t)$ with $t \in \real^{+}$ is
	completely monotonic {\rm{(CM)}} if it possesses derivatives $f^{(n)}(t)$
	for all $n=0,1,2,\dots$ and if these derivatives have alternating
	signs, that is, if 
	\begin{equation}
		(-1)^nf^{(n)}(t)\geq 0,\,\,\,\,t>0.
	\end{equation}
	The limit $f^{(n)}(0^{+})=\underset{t \rightarrow
0^+}{\lim} f^{(n)}(t)$, whether finite or infinite, must exist.
\end{defi}
\begin{defi}
A real non-negative function $g(t)$, defined in $[0,\infty)$, is called a
	Bernstein function if it possesses derivatives $g^{(n)}(t)$ for
	every $n=0,1,2,\dots$ and  
\begin{equation}
(-1)^ng^{(n)}(t)\leq0,\,\,\,\,t>0\,\,\,\,\mbox{for}\,\,\,n\in \nat ^*.
\end{equation}
\end{defi}

The first derivative of a Bernstein function is a CM function; examples of
Bernstein functions are $\phi(t)=1-e^{-t}$ and $\xi(t)=t^{\gamma}$, with
$0<\gamma\leq1$.

\begin{teo}

If $f(t)$, defined in $(0,\infty)$, is a  {\rm{CM}} function with 
	$0<f(t)\leq1$ for $t>0$, then the function $g(t)$, defined in 
	$[0,\infty)$ by equations 

\begin{eqnarray}
\begin{cases}
g(t):=1-f(t), & t>0 \\
g(t)=1, & t=0 
\end{cases}
\end{eqnarray}
is a Bernstein function.
\label{teofB}
\end{teo} 

An example of a Bernstein function is $g(t)=1-f(t)$ where $f(t)=e^{-t^\alpha}$
is the stretched exponential funcion and $\alpha \in (0,1]$, as
$f(t)=e^{-t^\alpha}$ is CM for $0<\alpha\leq1$. 

The following theorem is a fundamental result in the study of 
monotonicity as it provides a powerful tool to demonstrate the complete
monotonicity of many functions, including functions involving
Mittag-Leffler functions. 

\begin{teo}

A function $f(t)$ is {\rm{CM}} if and only if it can be represented as the
	Laplace transform {\rm{(LT)}} of a {\rm{(}}generalized{\rm{)}}
	non-negative function, i.e., if and only if 
\begin{equation}
f(t)=\mathscr{L}[K(r)](t)=\int_0^{\infty}e^{-rt}K(r)dr,\,\,\,\,\,K(r)\geq0,\,\,\,t\geq0,
\label{distri}
\end{equation}
where $K(r)$ is called spectral distribution function. 
\label{teoremaimportantee}
\end{teo}

This important theorem, called \textit{Bernstein's theorem}, is widely used
to demonstrate the complete monotonicity of functions. As an example we may
cite a work by Pollard \cite{Pollard:1948} in which it is demonstrated that
the Mittag-Leffler function $E_\alpha(-t)$ is {\rm{CM}} for
$0<\alpha\leq1$. In the case $\alpha =0$ this function is given by $E_0(-t)
= 1/(1+t)$, which is CM for $0<t<1$. 

In 1996, Schneider defined a generalized Mittag-Leffler function as
\cite{Schneider:1996}:

\begin{equation}
F_{\alpha,\beta}(t)=\Gamma(\beta)\cdot E_{\alpha,\beta}(-t),
\label{MLG}
\end{equation}
and demonstrated the complete monotonicity of this function if $\alpha$ and
$\beta$ satisfy 
\begin{equation}
0<\alpha\leq1\,\,\,\,\,\,\,\mbox{and}\,\,\,\,\,\,\alpha\leq\beta.
\end{equation}

Together with Bernstein's theorem, Titchmarsh's inversion formula
\cite{Titchmarsh:1948} constitutes a complete tool to study the
monotonicity of several types of functions. According to this formula, we
have the following relation: 

\begin{equation}
K(r)=-\frac{1}{\pi} \mbox{Im}\left\{h(s)\begin{large}|
\end{large}_{s=r\exp( i \pi)}\right\},
\label{imag}
\end{equation}
where 
\begin{equation}
h(s)= \mathscr{L}[f(t)](s).
\label{artigo26}
\end{equation}

With this result it has been possible to study the complete monotonicity of the
particular case in which the Kilbas-Saigo function turns into the particular
stretched exponential obtained by choosing $\alpha=1$  \cite{Capelas:2014}, as
follows: 

\begin{equation}
E_{1,1+\beta,\beta}(-t^{1+\beta})=\displaystyle \sum_{n=0}^{\infty}\displaystyle \prod_{i=0}^{n-1} \frac{\Gamma(i(1+\beta)+1+\beta)}{\Gamma(i(1+\beta)+2+\beta)}(-t^{\beta +1})^{n} =\displaystyle \sum_{n=0}^{\infty} (-1)^n t^{n(\beta+1)}\displaystyle \prod_{i=0}^{n-1} \frac{1}{(i+1)(\beta+1)},
\end{equation}
that is, 
\begin{equation}
E_{1,1+\beta,\beta}(-t^{1+\beta})=\displaystyle \sum_{n=0}^{\infty}\frac{1}{n!}\left(-\frac{t^{\beta+1}}{\beta+1}\right)^n=\exp\left(-\frac{t^{\beta+1}}{\beta+1}\right).
\end{equation}
Writing $\gamma=1+\beta$ we obtain the spectral distribution
function 
\begin{equation}
K_{1,\gamma}(r)=\frac{1}{r}\displaystyle\sum_{n=1}^{\infty}\frac{(-1)^n}{n!}\frac{(r^{-\gamma}/\gamma)^n}{\Gamma(-\gamma
n)}=\frac{1}{r}W_{-\gamma,0}(-r^{-\gamma}/\gamma),\,\,\,\,\,\,\,r>0,
\end{equation} 
where $W$ is a Wright function of the second type \cite{Mainardi:2010}.
Thus, this function is CM if, and only if, $0<\gamma<1$, or, equivalently,
$-1<\beta<0$.

With the help of the above results, it has been possible to prove the
complete monotonicity of several functions, among which functions that
generalize the functions studied by Pollard and Schneider. These functions
are defined in terms of Mittag-Leffler functions, which emerge in the study
of fractional differential equations. 

\section{Three-parameter Mittag-Leffler Function} 

The Mittag-Leffler function with three parameters is defined by
\cite{Prabhakar:1971}

\begin{equation}
E_{a,b}^{c}(z)=\sum_{k=0}^{\infty}\frac{(c)_k}{\Gamma(ak+b)}\frac{x^k}{k!},\,\,\,\,\,\,a,\,b,\,c
\in
\cplex\,\,\mbox{with}\,\,\mbox{Re}(a)>0,\,\mbox{Re}(b)>0,\,\mbox{Re}(c)>0,
\label{ml3p}
\end{equation}
where $(c)_k$ is the Pochhammer symbol, given by 
\begin{equation}
	(c)_k:=c(c+1)(c+2)\cdots(c+k-1)=\frac{\Gamma(c+k)}{\Gamma(c)},\,\,\,k\in\nat.
\end{equation}

Putting parameter $b=1$, we can introduce the function 
\begin{equation}
	\rho(t)=E_{\alpha,1}^{\lambda}(-t^{\alpha}),\,\,\,\,
\mbox{with}\,\,t\,\geq 0, \label{artigo1}
\end{equation}
with $\mbox{Re}(\alpha)>0$ and $\mbox{Re}(\lambda)>0$.
	
We observe in this definition that variable $t$ now has $\alpha$ as its exponent the
first parameter of the Mittag-Leffler function. We may thus conduct a first investigation
about its complete monotonicity by means of its spectral distribution function. 

For this sake, we apply the LT to the function defined in
Eq.(\ref{artigo1}); writing $\tilde{\rho}(s)=\mathscr{L}[\rho(t)](s)$ we have:

\begin{equation}
	\tilde{\rho}(s)=\sum_{k=0}^{\infty}\frac{(-1)^{k}(\lambda)_k}{k!}
	\frac{\mathscr{L}( t^{\alpha k})}{\Gamma(\alpha k +
	1)}=s^{-1}\sum_{k=0}^{\infty}\binom{-\lambda}{k}\left( \frac{1}{s}\right)^{\alpha
	k}=\frac{s^{\alpha \lambda-1}}{(s^{\alpha} +1)^{\lambda}}.  \label{artigo2}
\end{equation} 
Then, using Titchmarsh's inversion formula we find 

\begin{equation}
K_{\alpha,1}^{\lambda}(r)=-\frac{1}{\pi}\mbox{Im}\left\{\tilde{\rho}(s)|_{s=re^{i\pi}}\right\}.
\label{kr}
\end{equation} 

Substituting the expression found in Eq.(\ref{artigo2}) into Eq.(\ref{kr}) we obtain

\begin{equation}
	K_{\alpha,1}^{\lambda}(r)=-\frac{1}{\pi}\mbox{Im}\left\{\frac{(re^{i\pi})^{\alpha
	\lambda-1}}{[(re^{i\pi})^{\alpha}+ 1]^{\lambda}}\right\}=-\frac{r^{\alpha
	\lambda-1}}{\pi}\mbox{Im}\left\{\frac{e^{i\pi(\alpha
	\lambda-1)}}{[(re^{i\pi})^{\alpha}+ 1]^{\lambda}}\right\}. 
\end{equation}

We can simplify the quotient in the last member of this expression and take its imaginary
part. We then have 

\begin{equation} K_{\alpha,1}^{\lambda}(r)=-\frac{r^{\alpha
	\lambda-1}}{\pi}\mbox{Im}\left\{\frac{e^{i\pi(\alpha \lambda-1)}[r^{\alpha}
	e^{-i\alpha\pi}+ 1]^{\lambda}}{[1+2r^{\alpha}\cos(\alpha \pi)+ r^{2
	\alpha}]^{\lambda}}\right\} =\frac{r^{\alpha \lambda-1}}{\pi}
\frac{\sin[\theta \lambda+\pi(1-\alpha \lambda)]}{[1+2r^{\alpha}\cos(\alpha\pi)+ r^{2\alpha}]^{\lambda /2}},
\label{artigo59}
\end{equation}
where 
\begin{equation}
	\theta=\arctan \left[
	\frac{r^{\alpha}\sin(\alpha \pi)}{r^{\alpha}\cos(\alpha\pi)+1} \right],
	\,\,\,\,\,\, 0\leq\theta\leq\pi. 
\end{equation}

Denoting $K=K_{\alpha,1}^{\lambda}(r)$, we have from Eq.(\ref{artigo59}) that 
\begin{equation} K=\frac{r^{\alpha
\lambda-1}}{\pi}\frac{\sin[\theta \lambda+\pi(1-\alpha
\lambda)]}{[1+2r^{\alpha}\cos(\alpha\pi)+ r^{2\alpha}]^{\lambda
/2}}.  \label{artigo60}
\end{equation}
	
\begin{figure}[H]
\begin{center}
\includegraphics[scale=1.2]{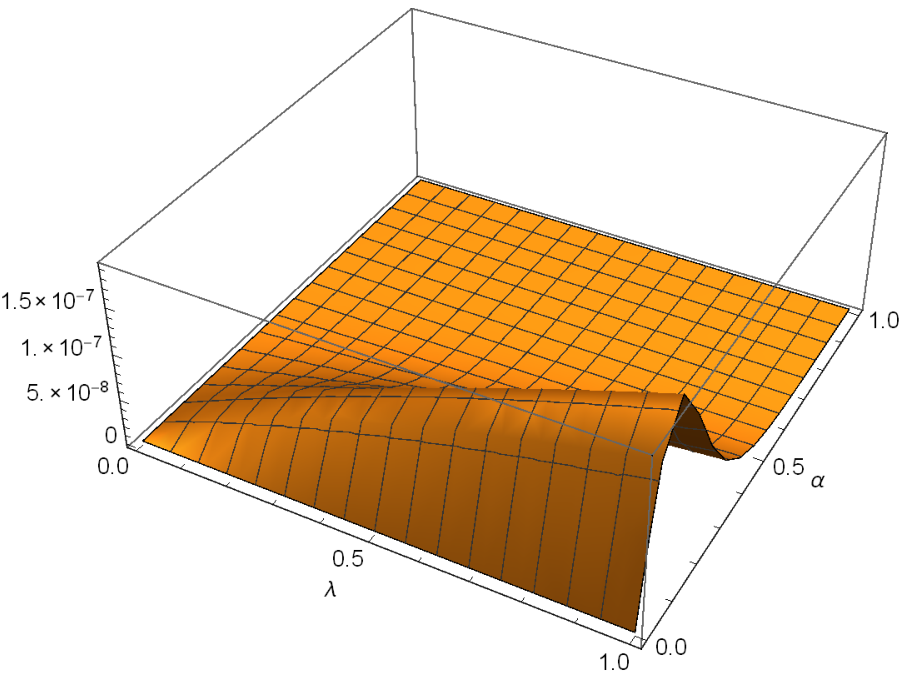} \qquad
\includegraphics[scale=1.2]{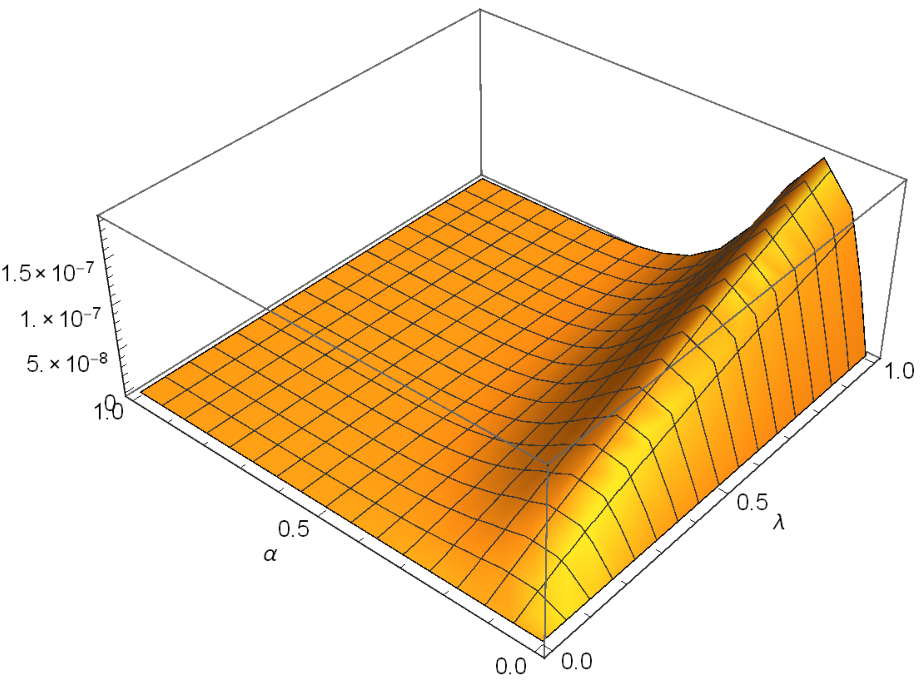}
\caption{Spectral distribution function 
$\rho(t)$, Eq.(\ref{artigo60}).}
\label{casopare}
\end{center}
\end{figure}

Observing the graphical behavior of distribution function $K$, for fixed $r$,
in Figure \ref{casopare}, we can conclude that as this distribution function is
non-negative for $0<\alpha\leq1$ and $0<\lambda\leq1$, the function $\rho(t)$
associated with it is CM when its parameters vary within these limits. 

\begin{figure}[H] \center
		\subfigure[ref1][$\lambda=0.5$]{\includegraphics[width=8.5cm]{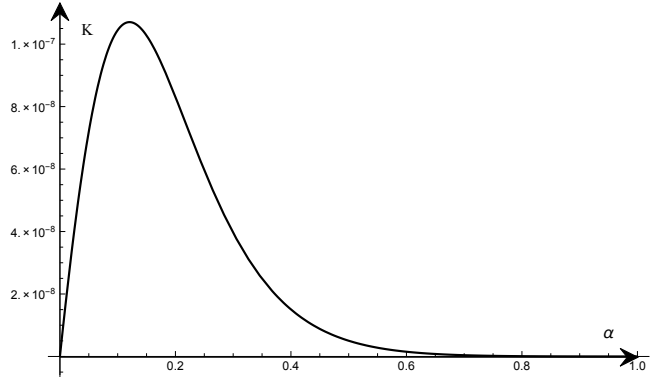}}
		\qquad
		\subfigure[ref2][$\alpha=0.5$]{\includegraphics[width=8.5cm]{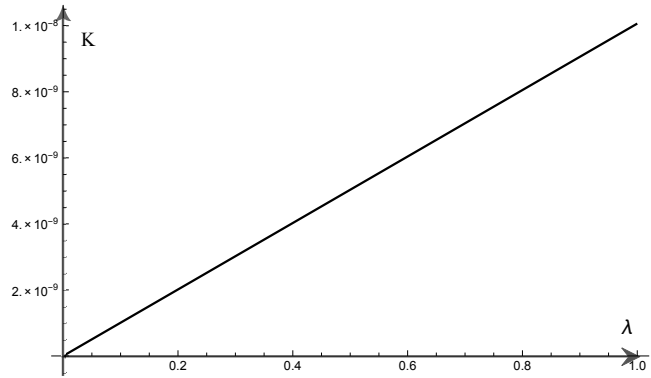}}
		\caption{Spectral distribution function $\rho(t)$,
		Eq.(\ref{artigo60}), with a fix parameter.} \label{casosparame}
\end{figure}

Furthermore, fixing one or the other of the remaining parameter, we obtain the graphics of
distribution function $K$ in $\real^{2}$, as can be seen in Figure \ref{casosparame}. In all
these graphics we considered a fixed value for $r$ in order to analyse the function's behavior
when we vary the values of its parameters. 

We use a theorem to demonstrate what we observe in the graphics. 

\begin{teo}
	
If $f(t)$ is a locally integrable function in $\real^{+}$ and {\rm{CM}} in $(0,\,\infty)$,
then its {\rm{LT}}, given by $\mathscr{L}[f(t)](s)=F(s)$, has the following properties: \\

\textbf{i)}
F(s) has an analytic extension in the region $\cplex \setminus \real^{-}$;\\
\textbf{ii)} $F(x) =  F^{*}(x)$ for $x \in (0,\infty)$;\\
\textbf{iii)} $\underset{x \rightarrow \infty}{\lim} F(x)=0$;\\
\textbf{iv)} $\mbox{Im}\{F(s)\}< 0$ for $\mbox{Im}\{s\} > 0$;\\
\textbf{v)} $\mbox{Im}\{sF(s)\}\geq0$ for $\mbox{Im}\{s\} > 0$ and $F(x)\geq 0$ for $x\in (0,
	\infty)$.
	
Reciprocally, every function $F(s)$ satisfying properties (i)-(iii), together with (iv) or (v),
	is the  {\rm{LT}} of a function $f(t)$ locally integrable in $\real^{+}$  and {\rm{CM}}
	in $(0,\,\infty)$.  \label{teoremaimportante}
\end{teo}
	
Indeed, we have from Eq.(\ref{artigo2}) that the LT of our function, 
	$\tilde{\rho}(s)=\mathscr{L}[E_{\alpha,1}^{\lambda}(-t^{\alpha})](s)$
	is given by:
\begin{equation} \tilde{\rho}(s)=
\frac{s^{\alpha\lambda-1}}{(s^{\alpha}+1)^\lambda}.
\label{laplacexi}
\end{equation} 
Hence, conditions ({\emph{i}})-({\emph{iii}}) are evidently satisfied.

Thus, as 
\begin{equation}
s\tilde{\rho}(s)=\frac{s^{\alpha\lambda}}{(s^{\alpha}+1)^{\lambda}}=\frac{1}{(1+s^{-\alpha})^{\lambda}},
\end{equation} 
we have to show that condition ({\emph{v}}) is also satisfied, that is, that for Im$\{s\} > 0$
one has $\mbox{Im}\{sF(s)\}\geq 0$.
	
Let us then write $s=re^{i\phi}$, with $0<\phi\leq \pi$; we get 
\begin{equation}
s\tilde{\rho}(s)=\frac{1}{(1+(re^{i
\phi})^{-\alpha})^{\lambda}}=\frac{(1+r^{-\alpha}e^{i\phi\alpha})^{\lambda}}{|1+r^{-\alpha}e^{-i\phi
\alpha}|^{2\lambda}}. 
\end{equation}
Now, putting $\Omega=|1+(re^{i\phi})^{-\alpha}|^{2\lambda}>0$, we have
\begin{equation}
\mbox{Im}\{s\tilde{\rho}(s)\}=\frac{1}{\Omega}\mbox{Im}\{(1+r^{-\alpha}e^{i\phi\alpha})^{\lambda}\}.
\end{equation}
		
Writing $z:=r^{-\alpha}e^{i\alpha\phi}$ and recalling that $0<\alpha\phi<\pi$, we conclude that
$z$ is in the first or second quadrant of the complex plane ($z \in \cplex^{+}$). Hence, adding
to $z$ the complex number $u=1$, whose argument is null, we obtain a new complex number which is
also in $\cplex^{+}$. Thus, $(1+z)^\lambda \in \cplex^{+}$, that is, Im$\{sF(s)\} \geq 0$, as we
wanted to demonstrate. 

Moreover, as $E_{\alpha,1}^{\lambda}(-t^\alpha)$ is CM for $0<\alpha\leq1$ and $0<\lambda\leq1$,
we have, according to Theorem (\ref{teofB}), that $f(t)=1-E_{\alpha,1}^{\lambda}(-t^\alpha)$ is a
Bernstein function.

\section{Kinetic Relaxation Equations}
		
The classical empirical laws associated with Debye (D), Cole-Cole (C-C),
Cole-Davidson (C-D) and Havriliak-Negami (H-N) models are, respectively, 

\begin{eqnarray}
\tilde{\varepsilon}_{D}(s)&=&\frac{1}{1+\sigma
s},\label{debye}\\
\tilde{\varepsilon}_{CC}(s)&=&\frac{1}{1+(\sigma
s)^\alpha},\label{coco}\\
\tilde{\varepsilon}_{CD}(s)&=&\frac{1}{(1+\sigma
s)^\beta},\label{coda}\\
\tilde{\varepsilon}_{HN}(s)&=&\frac{1}{(1+(\sigma
s)^\alpha)^\beta}\label{hn},
\end{eqnarray}
where $\alpha,\,\beta,\,\sigma\,\in\,\real$ and $\alpha,\,\beta\,\in\,\mbox{(}0,1\mbox{]}$.

In the study of relaxation processes in dielectrics it has been discovered, in
investigating the approximate linear response function, that the fluctuations
in polarization caused by thermal motion are equal to the fluctuations of the
macroscopic dipole relaxation function induced by electric fields
\cite{Williams:1972}. Thus, the laws governing the dipole
correlation function $\phi(t)$ are directly linked to the macroscopic
structural and kinetic properties of a dielectric system, which are represented
by function $\varphi (t)$. We may thus equate the relaxation function
$\varphi(t)$ and the macroscopic dipole correlation function $\phi(t)$, as
follows:
		
\begin{equation}
\varphi(t)\simeq\phi(t)=\frac{\langle
\mathcal{M}(t)\mathcal{M}(0)\rangle}{\langle \mathcal{M}(0)\mathcal{M}(0)\rangle},
\end{equation}
		
where $\mathcal{M}(t)$ is the macroscopic fluctuating dipole moment.
The dipole correlation function defined above is a particular case of the time
correlation function studied in the previous section. 
		
From the viewpoint of the modern theory of projection operators developed
by 
Mori \cite{Mori:1965} and Zwanzig \cite{Zwanzig:1961}, we have the
following integrodifferential equation for the time correlation function 
$\phi(t)$ associated with this memory function $\kappa(t)$,
known as \textit{memory function equation} \cite{Boon:1980}:
		
\begin{equation}
\frac{d\phi(t)}{dt}=-\int_{0}^{t}\kappa(x)\phi(t-x)dx,
\label{funcaoautocorrelacao}
\end{equation}
		
Introducing the concept of integral memory function, given by
$M(t)=\int_0^t \kappa(x)dx$ and using the fact that relaxation function
$\varphi(t)$ also satisfies Eq.(\ref{funcaoautocorrelacao}), we obtain the
following relation for function $\varphi(t)$:
		
\begin{equation}
\frac{d\varphi(t)}{dt}=-\frac{d}{dt}\int_0^tM(t-x)\varphi(x)dx\equiv-\frac{d}{dt}\left[M(t)\ast\varphi(t)\right],
\label{convolucao}
\end{equation}
		
where $\ast$ denotes convolution product. 
		
We also know that there exists between relaxation function $\varphi(t)$ and
dielectric constant $\tilde{\varepsilon}(s)$ the following relation
involving LT: 
		
\begin{equation}
\tilde{\varepsilon}(s)=\int_0^{\infty}\left(-\frac{d\varphi}{dt}\right)
e^{-s t}dt=\mathscr{L}\left[-\frac{d\varphi(t)}{dt}\right](s).
\label{permiss}
\end{equation}
		
Thus, from Eq.(\ref{convolucao}), Eq.(\ref{permiss}) and the empirical laws
given by Eqs.(\ref{debye})-(\ref{hn}) we obtain the following kinetic
relaxation equations: 

\begin{eqnarray}
\mbox{D}& &\frac{d\varphi(t)}{dt}+\frac{1}{\sigma}\varphi(t)=0 ; \label{eqcin1}\\
\mbox{C-C}& &\frac{d\varphi(t)}{dt}+\frac{1}{\sigma^{\alpha}}D_{t}^{1-\alpha}\varphi(t)=0 ; \\
\mbox{C-D}&&\frac{d\varphi(t)}{dt}+\frac{1}{\sigma^{\beta}}\frac{d}{dt}\left\{e^{-t/\sigma}\int_0^t(t-x)^{\beta-1}E_{\beta,\beta}^{1}\left[\left(\frac{t-x}{\sigma}\right)^{\beta}\right]e^{x/\sigma}\varphi(x)dx\right\}=0 ; \label{eqcin3}\\
\mbox{H-N}&
&\frac{d}{dt}\left\{\varphi(t)+
\sum_{k=0}^{\infty}\int_0^t\frac{(t-x)^{\alpha\beta(k+1)-1}}{\sigma^{\alpha\beta(k+1)}}
E_{\alpha,\alpha\beta(k+1)}^{\beta(k+1)}\left[-\left(\frac{t-x}{\sigma}\right)^{\alpha}\right]\varphi(x)dx\right\}=0 . \label{eqcin4}
\end{eqnarray}
Here, $E_{a,b}^{c}(\cdot)$ are Mittag-Leffler functions and $D_{t}^{\gamma}f(t)$ denotes the fractional differential in the 
Riemann-Liouville sense, defined by 
		
\begin{equation}
D_{t}^{\gamma}f(t)=\frac{1}{\Gamma(1-\gamma)}\frac{d}{dt}\int_0^t
\frac{f(x)}{(t-x)^{\gamma}}dx,\,\,\,\,\,\,\,0<\gamma\leq1.
\label{riemann}
\end{equation}
		
Let us now consider Eq.(\ref{eqcin4}), associated with the H-N model, in the
following form: 

\begin{eqnarray}
\frac{d}{dt}\varphi_{HN}(t)=-\frac{d}{dt}\left\{\left[\sum_{k}\frac{t^{\alpha\beta(k+1)-1}}{\sigma^{\alpha\beta(k+1)}}E_{\alpha,\alpha\beta(k+1)}^{\beta(k+1)}\left[-\left(\frac{t}{\sigma}\right)^{\alpha}\right]\right]*\varphi_{HN}(t)\right\}.
\label{hnforma}
\end{eqnarray}

Applying LT to Eq.(\ref{hnforma}) we get 
\begin{eqnarray}
s\tilde{\varphi}_{HN}(s)-1=-s\sum_{k}\frac{1}{(1+(s\sigma)^{\alpha})^{\beta(k+1)}}\cdot\tilde{\varphi}_{HN}(s),
\end{eqnarray}
where we considered $\varphi_{HN}(0)=1$.

This can be rewritten as 
\begin{eqnarray}
s\tilde{\varphi}_{HN}(s)-1=-s\frac{\tilde{\varphi}_{HN}(s)}{(1+(s\sigma)^{\alpha})^{\beta}}\sum_{k}
\frac{1}{(1+(s\sigma)^{\alpha})^{\beta k}}.
\end{eqnarray}
Then, calculating the sum on the right-hand side, it follows that 
\begin{eqnarray}
s\tilde{\varphi}_{HN}(s)-1=-s\frac{\tilde{\varphi}_{HN}(s)}{(1+(s\sigma)^{\alpha})^{\beta}}
\frac{1}{1-(1+(s\sigma)^{\alpha})^{-\beta}}.
\end{eqnarray}
Isolating $\tilde{\varphi}_{HN}(s)$ we get 
\begin{eqnarray}
\tilde{\varphi}_{HN}(s)=\frac{1}{s}-\frac{1}{\sigma^{-\alpha \beta}}\frac{s^{-1}}{(s^{\alpha}+\sigma^{-\alpha})^{\beta}}.
\end{eqnarray}
Applying the inverse LT we obtain the solution 
\begin{eqnarray}
\varphi_{HN}(t)=1-\left(\frac{t}{\sigma}\right)^{\alpha\beta}E_{\alpha,\alpha\beta+1}^{\beta}\left[-
\left(\frac{t}{\sigma}\right)^{\alpha}\right].
\label{hncap3}
\end{eqnarray}

Applying the same procedure to Eqs.(\ref{eqcin1})-(\ref{eqcin3}) we obtain
their respective solutions:

\begin{eqnarray}
\mbox{D}&\,\,&\varphi_{D}(t)=e^{-t/\sigma},\label{debye0}\\
\mbox{CC}&\,\,&\varphi_{CC}(t)=E_{\alpha}\left[-\left(\frac{t}{\sigma}\right)^{\alpha}\right],\label{cc}\\
\mbox{CD}&\,\,&\varphi_{CD}(t)=1-\left(\frac{t}{\sigma}\right)^{\beta}E_{1,\beta+1}^{\beta}\left[-\frac{t}{\sigma}\right].\label{cd}
\end{eqnarray}

\begin{figure}[H]
\begin{center}
\includegraphics[width=7.5cm]{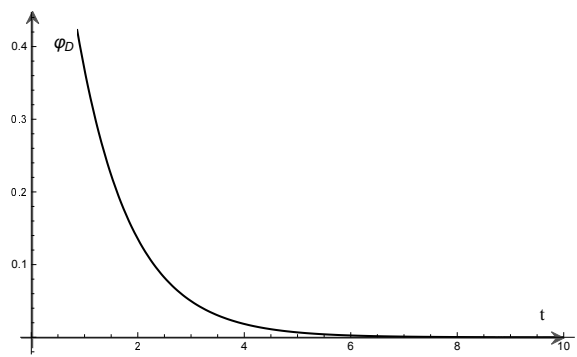}
\caption{Debye function.}
\label{naovai}
\end{center}
\end{figure}

\begin{figure}[H]
\subfigure[$\alpha=0.4$]{\includegraphics[width=7.5cm]{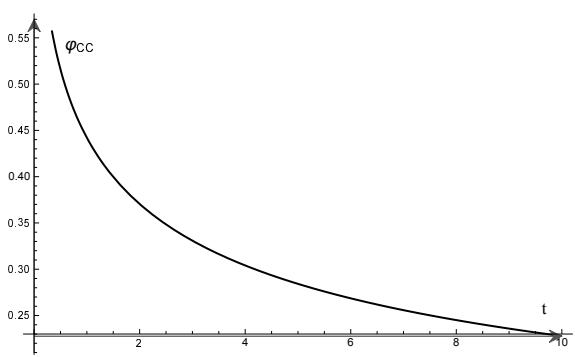}}
\qquad
\subfigure[ref2][$\alpha=0.6$]{\includegraphics[width=7.5cm]{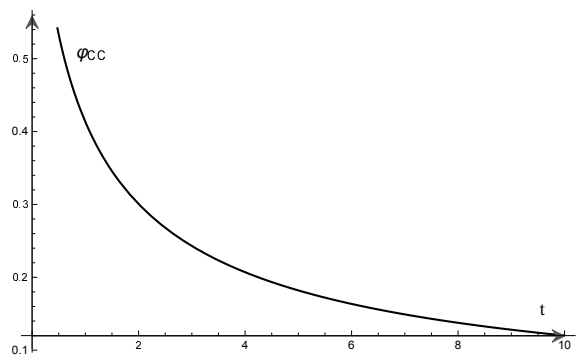}}
\caption{Cole-Cole function $\varphi_{CC}(t)=E_{\alpha}(-t^{\alpha})$.}
\label{coole}
\end{figure}

\begin{figure}[H]
\center
\subfigure[$\beta=0.4$]{\includegraphics[width=7.5cm]{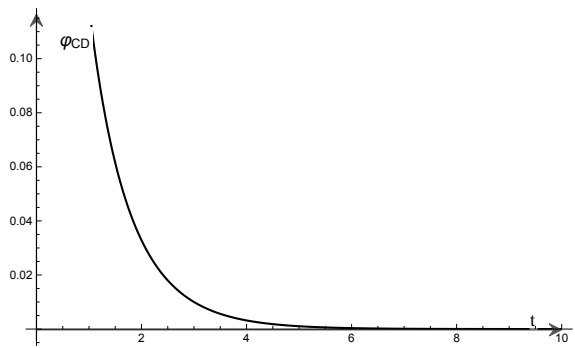}}
\qquad
\subfigure[ref2][$\beta=0.6$]{\includegraphics[width=7.5cm]{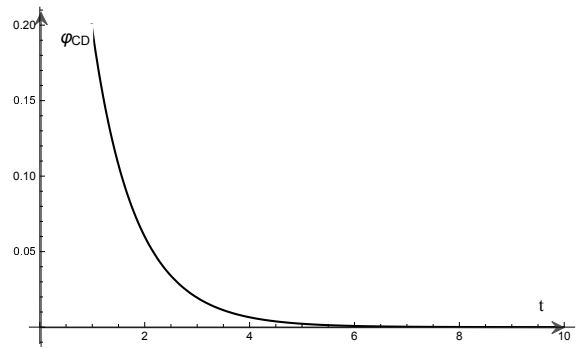}}
\caption{Cole-Davidson function $\varphi_{CD}(t)=1-t^{\beta}\times E_{1,\beta+1}^{\beta}(-t)$ with $0<\beta<1$.}
\label{daavi}
\end{figure}

We can easily see that if we consider $\alpha =1$ in the solution given by 
Eq.(\ref{hncap3}) we recover the solution given by Eq.(\ref{cd}). On the
other hand, if we take  
$\beta=1$ in the solution given by Eq.(\ref{hncap3}), 
we recover Eq.(\ref{cc}). Indeed,

\begin{eqnarray}
\varphi_{HN}(t)|_{\beta=1}=1-\left(\frac{t}{\sigma}\right)^{\alpha}E_{\alpha,\alpha+1}\left[-\left(\frac{t}{\sigma}\right)^{\alpha}\right]=1+\sum_{k}\frac{\left[-\left(\frac{t}{\sigma}\right)^{\alpha}\right]^{k+1}}{\Gamma(\alpha(k+1)+1)}.
\label{artigo15}
\end{eqnarray}
Writing $n=k+1$ in the summation in the second member
of Eq.(\ref{artigo15}) we finally get
\begin{eqnarray}
\varphi_{HN}(t)|_{\beta=1}= \sum_{n}\frac{\left[-\left(\frac{t}{\sigma}\right)^{\alpha}\right]^{n}}{\Gamma(\alpha n+1)}=\varphi_{CC}(t).
\end{eqnarray}
Thus, if we consider $\alpha = \beta = 1$ in Eq.(\ref{hncap3}) we recover the
solution given by Eq.(\ref{debye0}).  Graphic representations of the solutions
given by Eq.(\ref{debye0}), Eq.(\ref{cc}) and Eq.(\ref{cd}) are shown in
Figures \ref{naovai}, \ref{coole} and \ref{daavi}, respectively. 


\section{Kinetic Relaxation Functions}

In order to make easier the demonstration of the complete monotonicity of
these functions, we assume, without loss of generality, that $\sigma =1$.
Besides, we must keep in mind that the parameters $\alpha$ and $\beta$
appearing in kinetic relaxation equations belong to the real interval
$(0,1]$. 

\begin{figure}[H]
\begin{center}
\includegraphics[width=20cm]{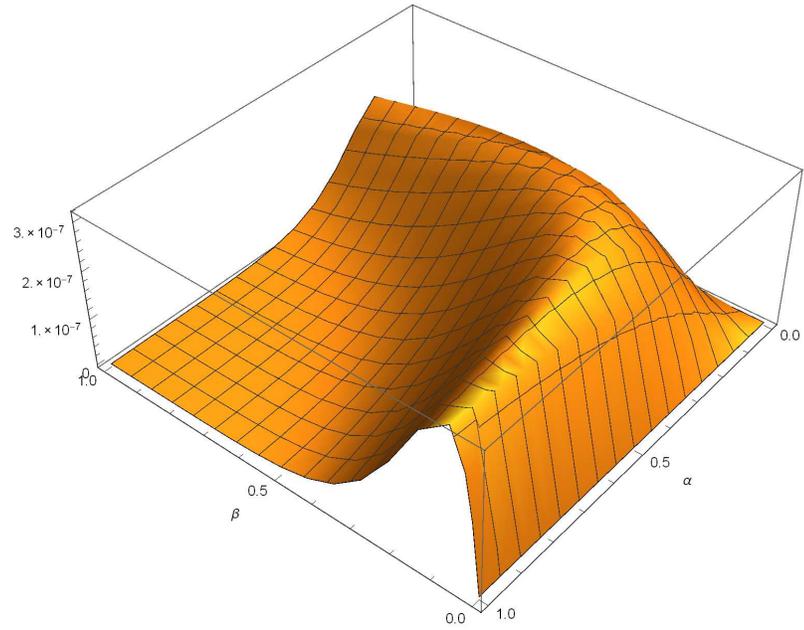}
\end{center}
\caption{Spectral distribution function H-N according to Eq.(\ref{artigo12}).}
\label{artigo4}
\end{figure}

\begin{figure}[H]
\center
\subfigure[ref1][$\beta=0.3$]{\includegraphics[width=7.5cm]{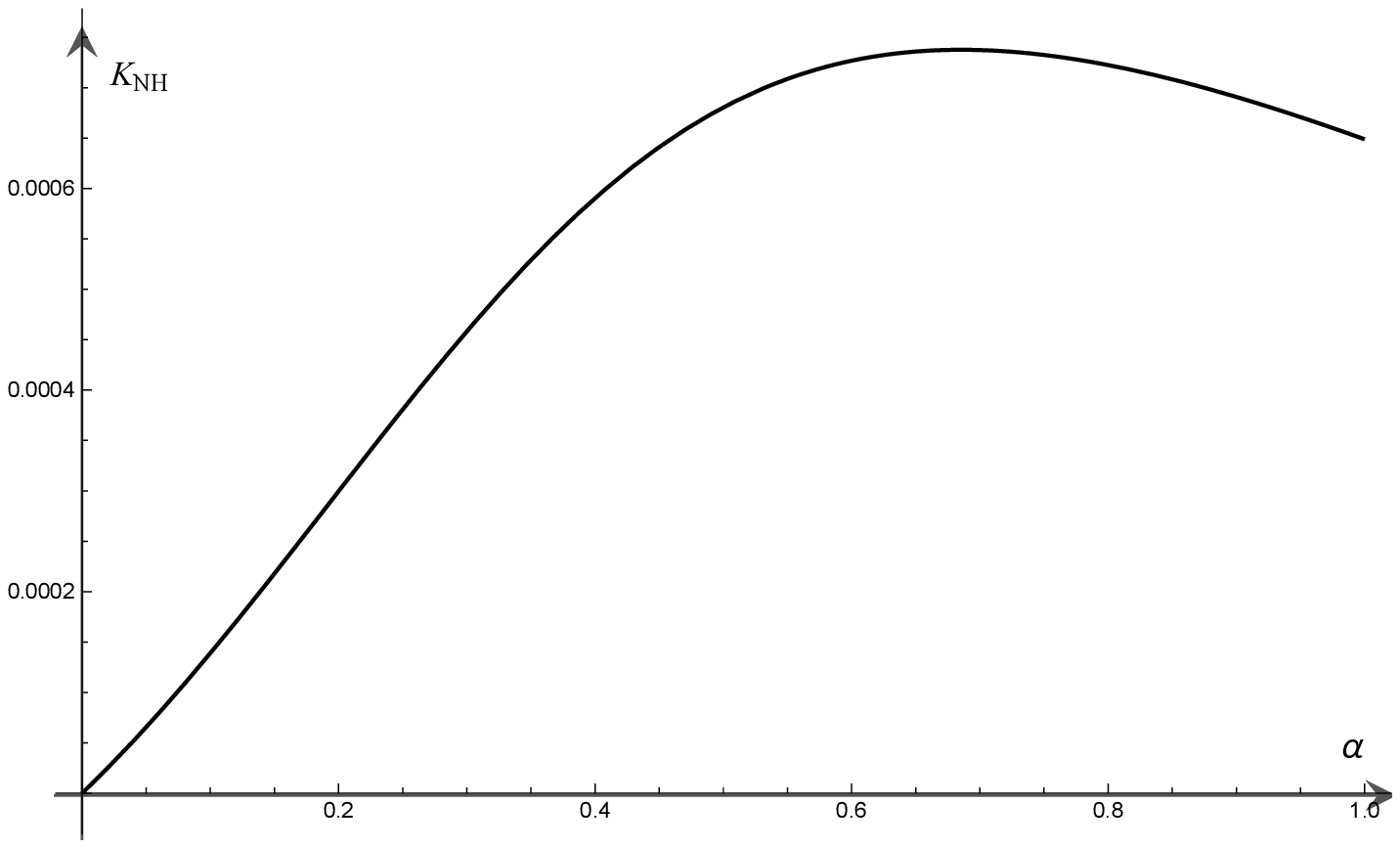}}
\qquad
\subfigure[ref2][$\beta=0.8$]{\includegraphics[width=7.5cm]{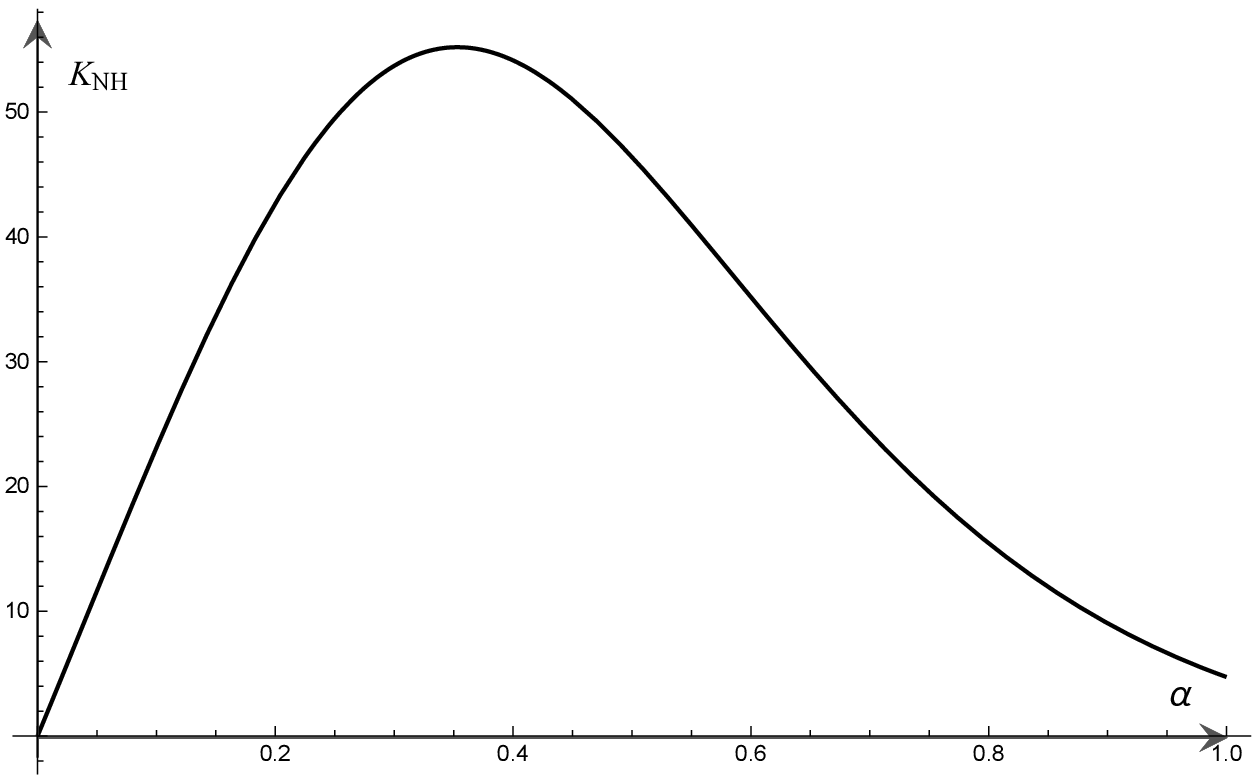}}
\caption{Spectral distribution function H-N for $0<\alpha\leq1$, according
	to Eq.(\ref{artigo12}).}
\label{artigo5}
\end{figure}

\begin{figure}[H]
\center
\subfigure[ref1][$\alpha=0.3$]{\includegraphics[width=7.5cm]{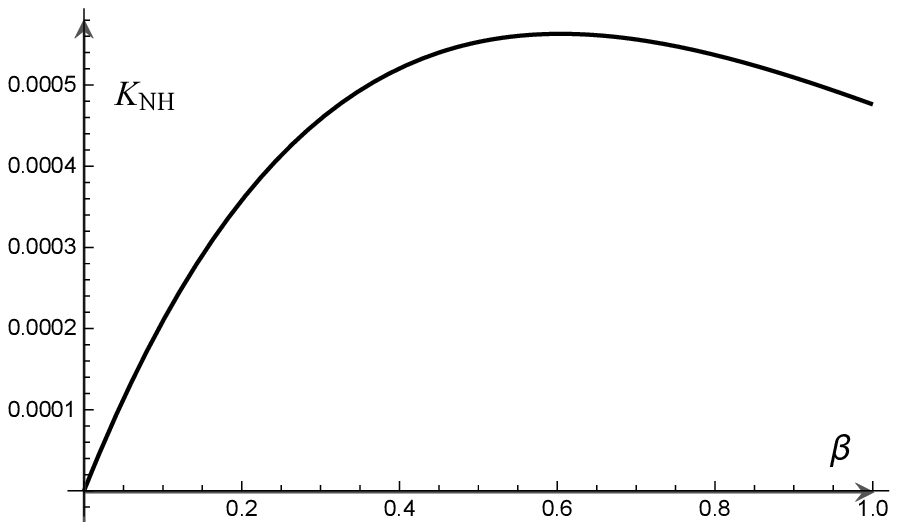}}
\qquad
\subfigure[ref2][$\alpha=0.8$]{\includegraphics[width=7.5cm]{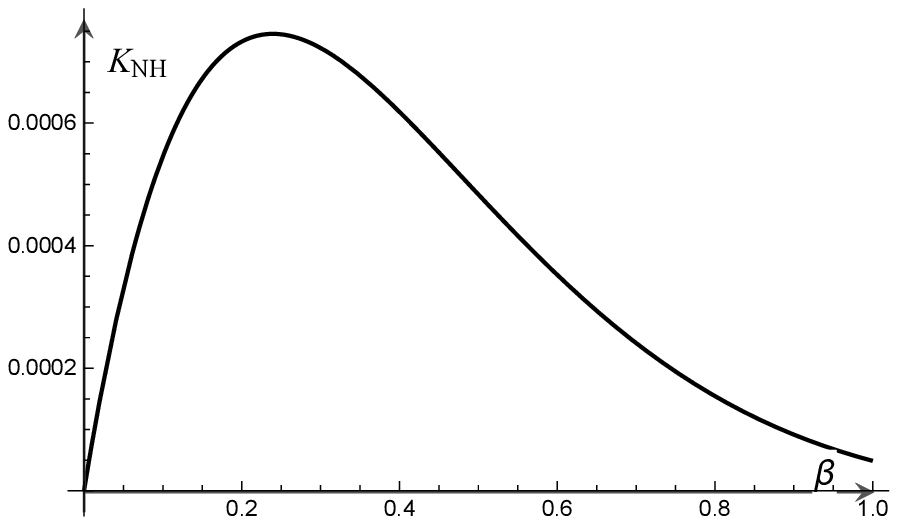}}
\caption{Spectral distribution function H-N for $0<\beta\leq1$, according to Eq.(\ref{artigo12}).}
\label{artigo6}
\end{figure}

\begin{itemize}

\item Beginning with the Debye function, given by $\varphi_{D}(t)=e^{-t}$,
	we can conclude from the very definition of monotonicity that this
		function is CM. Indeed, 

\begin{equation}
(-1)^{n}D^{n}[\varphi_{D}(t)]=(-1)^{2n}e^{-t}=e^{-t}=\varphi_{D}(t)>0,\,\,\,t\geq0.
\end{equation}
Besides, according to 
Eq.(\ref{distri}) of Theorem (\ref{teoremaimportantee}),
its spectral distribution function is the Dirac delta
function $\delta(r-1)$, that is, it is a positive function. 

\item Let us now consider C-C function which, as we have seen, is defined
	by 
\begin{equation}
\varphi_{CC}(t)=E_{\alpha}(-t^{\alpha}),\,\,\,\,\mbox{with}\,\,\,\,t\geq0.
\end{equation}
This function is identical to a Mittag-Leffler with a single parameter
		$\alpha$. On the other hand, this latter function is a
		particular case of the function given by Eq.(\ref{artigo1})
		for $\lambda=1$. In the previous section, we proved its
		complete monotonicity with the help of Titchmarsh's formula
		and Theorem (\ref{teoremaimportante}). We can thus conclude
		that function $\varphi_{CC}(t)$ is CM for $0<\alpha\leq1$.\\

\item Consider now the C-D function written in the following form: 
\begin{equation}
\varphi_{CD}(t)=1-t^{\beta}E_{1,\beta+1}^{\beta}(-t),\,\,\,\,
	\mbox{with}\,\,\,\,t\geq0.
\end{equation}
This function involves a Mittag-Leffler function with three parameters, one
		of which is equal to $1$ while the other two are written in
		terms of $\beta$. We can show that it is CM for $0 < \beta <
		1$. 
\begin{proof}

Indeed, let $\tilde{\varphi}_{CD}(s)=\mathscr{L}[\varphi_{CD}(t)](s)$, the
LT of function $\varphi_{CD}(t)$, given by 

\begin{equation}
\tilde{\varphi}_{CD}(s)=\frac{1}{s}\left[1-\frac{1}{(s+1)^\beta}\right].
\end{equation}
We know that its spectral distribution function has the form 
\begin{equation}
K_{\beta}(r)=\frac{1}{\pi}\mbox{Im}\left\{\tilde{\varphi}_{CD}(s)| _{s=re^{-i\pi}}\right\}
\end{equation}
This spectral distribution has a branch cut.  Hence, $K_{\beta}(r)=0$ when
$r<1$ and 
\begin{eqnarray}
K_{\beta}(r)=\frac{1}{\pi r}\frac{\sin(\beta\theta)}{(r-1)^{2\beta}},
	\,\,\,\,\,\mbox{for}\,r>1,
\label{artigo41}
\end{eqnarray}
with 
\begin{equation}
\theta= \arctan \left(\frac{r\sin\pi}{1+r\cos\pi}\right).
\end{equation}
Consider the complex number 
\begin{equation}
\rho e^{i \theta }=1+re^{i \pi }.
\label{artigo42}
\end{equation} 

If we consider $\theta=0$, it follows from the last equation that
	$\rho=1-r$.  This is not a solution because, according to
	Eq.(\ref{artigo41}), $r>1$ and so $\rho$ would be negative. 

On the other hand, if we choose $\theta =\pi$ in Eq.(\ref{artigo42}) we
	have $\rho=r-1$; as $r>1$, radius $\rho$ will have positive values. 

Hence, 
\begin{equation}
K_{\beta}(r)=\frac{1}{\pi r}\frac{\sin(\beta\pi)}{(r-1)^{2\beta}}.
\end{equation}
We conclude that in order to have $K_{\beta}(r)\geq0$, it is necessary and
	sufficient that $\beta \pi \in \cplex^+$, i.e., that $0<\beta\leq1$.
\end{proof}

\item Let us now consider the most general relaxation model and its
	function, H-N, written in the form 
\begin{equation}
\varphi_{HN}(t)=1-t^{\alpha\beta}
	E_{\alpha,\alpha \beta+1}^{\beta}(-t^{\alpha}),
	\,\,\,\,\mbox{with}\,\,\,\,t\geq0.
\label{hnfunction}
\end{equation}

We shall use Bernstein's theorem to analyse its complete monotonicity. Also,
a graphical analysis will provide some important insights. 


We begin calculating, with the help of Eq.(\ref{imag}), the spectral
distribution function associated with H-N. We find 
\begin{equation}
K_{HN}=\frac{\sin \left[\beta\vartheta\right]}{\pi r (1+2r^\alpha\cos(\alpha
	\pi)+r^{2\alpha})^{\beta/2}},
\label{artigo12}
\end{equation}
where $\vartheta=\vartheta(r,\alpha)$ is given by 
\begin{equation}
\vartheta=\arctan \left(\frac{r^\alpha \sin(\alpha \pi)}{1+r^{\alpha}\cos(\alpha\pi)}\right),\,\,\,\,0<\vartheta\leq \pi.
\end{equation}

Since $0<\vartheta\leq \pi$ and $0<\beta\leq1$, it follows that $0<\beta
\vartheta\leq\pi$. This implies that 
$\sin [\beta\vartheta]\geq 0$. Also, remark that the denominator in the
expression of $K_{HN}$ is a positive number and will not influence its sign. 
We thus conclude that the spectral distribution function $K_{HN}$
associated with the H-N model is everywhere positive. 

Figure \ref{artigo4} presents a plot in $\real^3$ of the spectral
distribution function in terms of variables $\alpha$ and $\beta$. It shows 
that this function is always positive for $0<\alpha\leq1$ and
$0<\beta\leq1$. In Figures \ref{artigo5} and \ref{artigo6} we fixed one
parameter in order to plot the distribution's behavior as a function of the
remaining parameter. In all plots the value of $r$ is held constant. 

\end{itemize}

\section{Fractional Kinetic Relaxation Functions} 

Let us now consider the generalization of the relaxation model by means of the
fractional differential equation given by 

\begin{equation}
D_{t}^{\gamma}\varphi(t)=-D_{t}^{\gamma}\left\{M(t)\ast\varphi(t)\right\},
	\,\,\,\,\mbox{with}\,\gamma \in (0,1], 
\label{fracionaria}
\end{equation}
where operator $D_{t}^{\gamma}$ is the fractional derivative in the
Riemann-Liouville sense, shown in Eq.(\ref{riemann}). 

In the fractional case, the complex dielectric permissivity (in variable $s$)
is given by the following superposition relation \cite{Frohlich:1958}:
\begin{equation}
\tilde{\varepsilon}(s)=\mathscr{L}\left[-\frac{d^{\gamma}\varphi(t)}{dt}\right](s).
\label{constfrac}
\end{equation}

Thus, from Eq.(\ref{fracionaria}), Eq.(\ref{constfrac}), the normalization
condition for relaxation function $\varphi(t)$, given by
\begin{equation}
D^{\gamma-1}_{t}\varphi(0)=1,
\label{norma}
\end{equation}
and the classical empirical laws given by Eqs.(\ref{debye})--(\ref{hn}), 
we obtain the memory functions associated with the models by D, C-C, C-D and
H-N: 
 \begin{eqnarray}
\mbox{D}&\,\,\,\,\,\,&M_{D}(t)=\frac{1}{\sigma},\label{MT1}\\
\mbox{C-C}&\,\,\,\,\,\,&M_{CC}(t)=\frac{t^{\alpha-1}}{\sigma^{\alpha}\Gamma(\alpha)},\label{MT2}\\
\mbox{C-D}&\,\,\,\,\,\,&M_{CD}(t)=e^{-t/\sigma}\frac{t^{\beta-1}}{\sigma^{\beta}}E_{\beta,\beta}\left[\left(\frac{t}{\sigma}\right)^{\beta}\right],\label{MT3}\\
\mbox{H-N}&\,\,\,\,\,\,&M_{HN}(t)=\sum_{k=0}^{\infty}\left(\frac{t}{\sigma}\right)^{\alpha\beta(k+1)}t^{-1}E_{\alpha,\alpha\beta(k+1)}^{\beta(k+1)}\left[-\left(\frac{t}{\sigma}\right)^{\alpha}\right],\label{MT4}
\end{eqnarray}
where $\alpha,\,\beta,\,\sigma\,\in\,\real$ and $\alpha,\,\beta\,\in\,\mbox{(}0,1\mbox{]}$.

From these memory functions we can obtain the fractional kinetic relaxation
equations. The solutions of such fractional kinetic relaxation equations are
called fractional kinetic relaxation functions. We shall now discuss their
monotonicity.


\begin{itemize}
\item The solution of the fractional differential equation 
\begin{equation}
D^{\gamma}_{t}\varphi_{DF}(t)=
- D^{\gamma}_{t}\left\{M_{D}\ast\varphi_{DF}(t)\right\}=
- D^{\gamma}_{t}\left\{\frac{1}{\sigma}\ast\varphi_{DF}(t)\right\}
\label{debyef1}
\end{equation}
is the fractional Debye function, given by 
\begin{equation}
\varphi_{DF}(t)= t^{\gamma-1}E_{1,\gamma}(-t),\,\,\,\,t\geq0,
\label{artigo9}
\end{equation}
with $0<\gamma\leq 1$. 

This function is the product of a power function and a
		Mittag-Leffler function with two parameters, with the
		exponent of the power function equal to the second
		parameter of the Mittag-Leffler function. Figure
		\ref{artigo11} shows the graphics of this fractional Debye
		function  for two values of $\gamma$.  Before analysing
		this function more thouroughly we shall consider the
		fractional relaxation function associated with the C-C
		model.

\begin{figure}[H]
\center
\subfigure[ref1][$\gamma=0.3$]{\includegraphics[width=7.5cm]{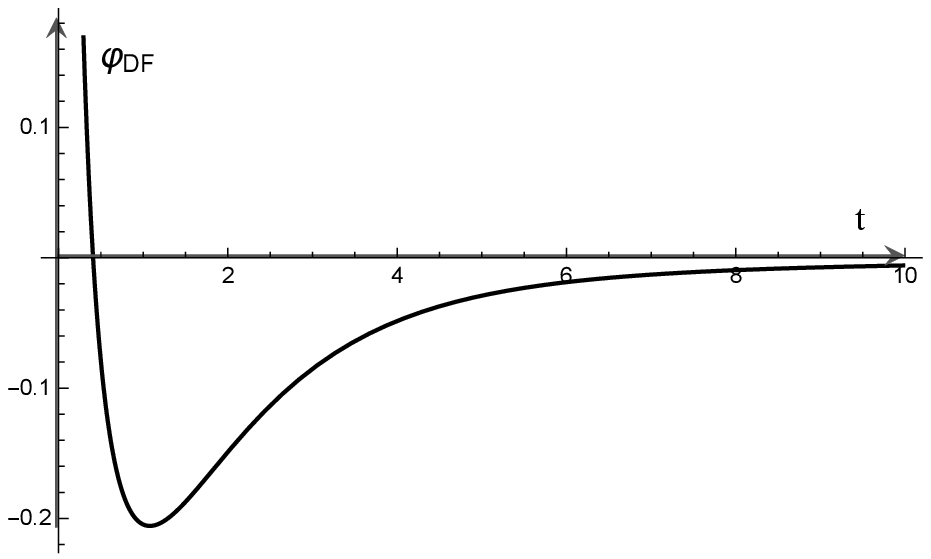}}
\qquad
\subfigure[ref4][$\gamma=0.9$]{\includegraphics[width=7.5cm]{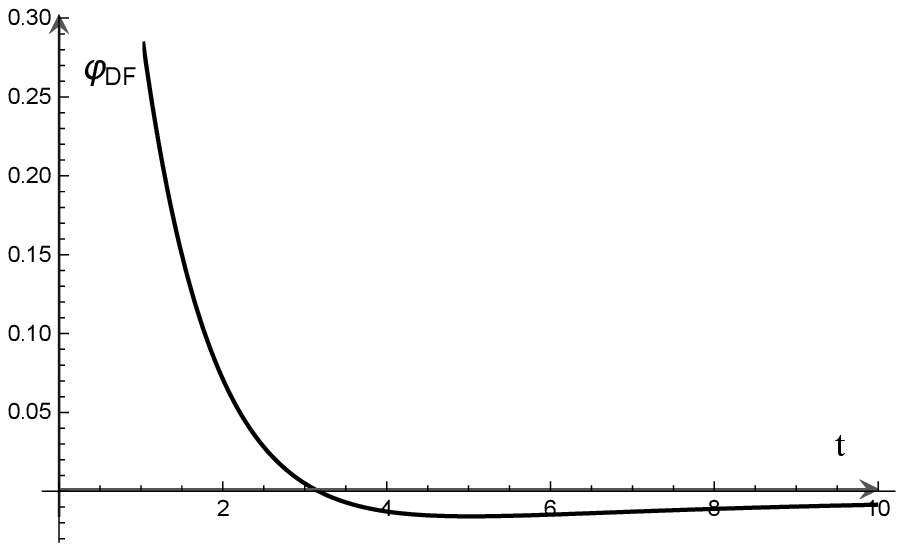}}
\caption{Function $\varphi_{DF}(t)= t^{\gamma-1}E_{1,\gamma}(-t)$.}
\label{artigo11}
\end{figure}
\item Starting with the C-C fractional kinetic relaxation equation,
\begin{equation}
D^{\gamma}_{t}\varphi_{CCF}(t)=- D^{\gamma}_{t}\left\{M_{CC}\ast\varphi_{CCF}(t)\right\}=- D^{\gamma}_{t}\left\{\frac{t^{\alpha-1}}{\sigma^{\alpha}\Gamma(\alpha)}\ast\varphi_{CCF}(t)\right\},
\label{ccf1}
\end{equation}
we obtain its solution, the fractional C-C function, 
\begin{equation}
\varphi_{CCF}(t)= t^{\gamma-1}E_{\alpha,\gamma}(-t^\alpha),
	\,\,\,\,\, t\geq0.
\label{artigo40}
\end{equation}
The LT of $\varphi_{CCF}(t)$, denoted by $\tilde{\varphi}_{CCF}(s)=
		\mathscr{L}[\varphi_{CCF}(t)](s)$, is given by 
\begin{equation}
\tilde{\varphi}_{CCF}(s)=\frac{s^{\alpha-\gamma}}{s^{\alpha}+1}
\end{equation}
Thus, its distribution function is 
\begin{equation}
K_{CCF}=\frac{r^{\alpha-\gamma}}{\pi}
	\frac{\sin[(\gamma-\alpha)\pi+\theta]}{[r^{2\alpha}+2r^{\alpha}\cos(\alpha \pi)+1]^{1/2}},
\label{artigo61}
\end{equation}
where 
\begin{equation}
\theta= \arctan\left[\frac{r^{\alpha}\sin(\alpha \pi)}{1+r^{\alpha}\cos(\alpha \pi)}\right].
\end{equation}
As $0 < \theta < \alpha \pi$, we have $-\alpha \pi < \theta - \alpha \pi <0$ and thus
$\gamma \pi - \alpha \pi < \theta + (\gamma - \alpha) \pi < \gamma \pi$. We impose
that $\alpha \leq \gamma$ and we know that $0 < \alpha, \, \gamma \leq 1$; we thus
have $0 \leq (\gamma -\alpha) \pi < \theta + (\gamma -\alpha) \pi < \gamma \pi \leq
\pi$, i.e. $0 < \theta + (\gamma -\alpha)\pi < \pi$. We thus conclude that, if $0 <
\alpha \leq \gamma \leq 1$, then the sine function in Eq.(\ref{artigo61}) is
non-negative and so $K_{CCF} \geq 0$. 

In Figures \ref{artigo32} and \ref{artigo33} we can see the graphic
behavior of the spectral distribution function associated with the
fractional C-C function. In particular, Figure \ref{artigo33} shows that
the spectral distribution function, though very close to zero, becomes
positive when $\gamma$ exceeds $\alpha$ (Figure \ref{artigo33}(a)); when
$\alpha$ exceeds $\gamma$ the function becomes negative (Figure
\ref{artigo33}(b)). 

In Figures \ref{artigo7} and \ref{artigo8} we can verify, for two different
fixed values of $\gamma$, the changes in the function's behavior according
to whether parameter $\alpha$ is smaller or greater than parameter
$\gamma$. In both figures, $\alpha < \gamma$ in (a) and (b) and $\alpha >
\gamma$ in (c) and (d). 

\begin{figure}[H]
\center
\includegraphics[width=11cm]{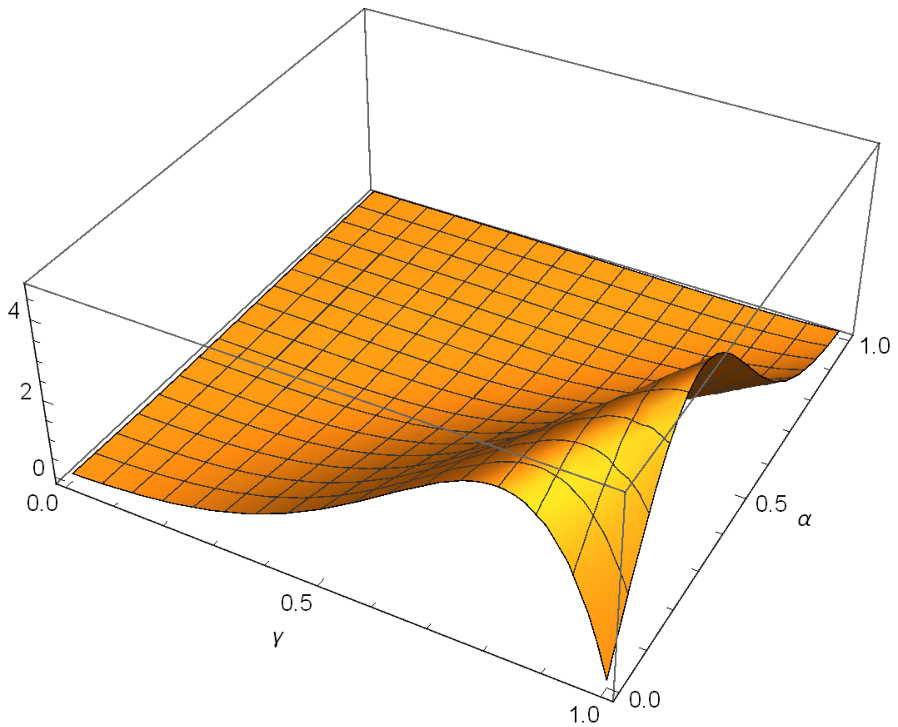}
\caption{Spectral distribution associated with fractional C-C function,
	Eq.(\ref{artigo61}).}
\label{artigo32}
\end{figure}
\begin{figure}[H]
\center
\subfigure[ref1][$\alpha=0.4$]{\includegraphics[width=7.5cm]{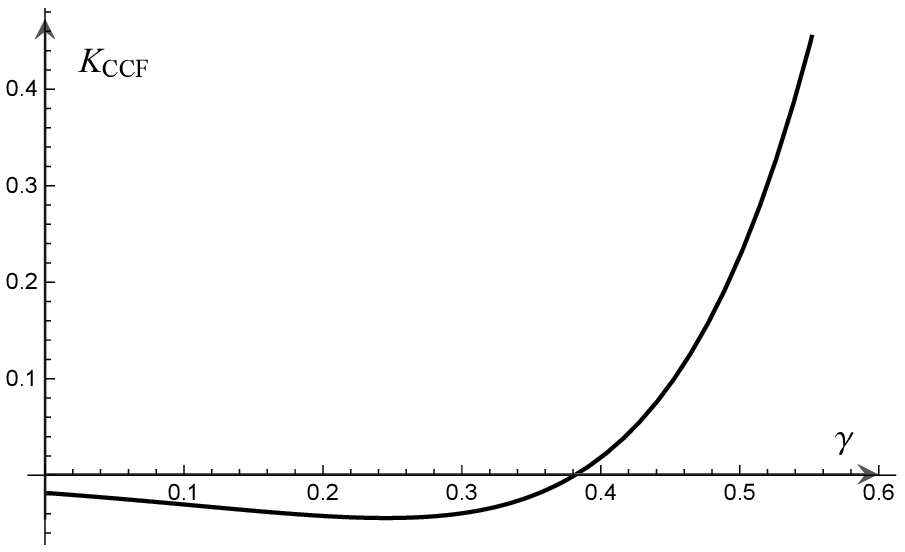}}
\qquad
\subfigure[ref4][$\gamma=0.6$]{\includegraphics[width=7.5cm]{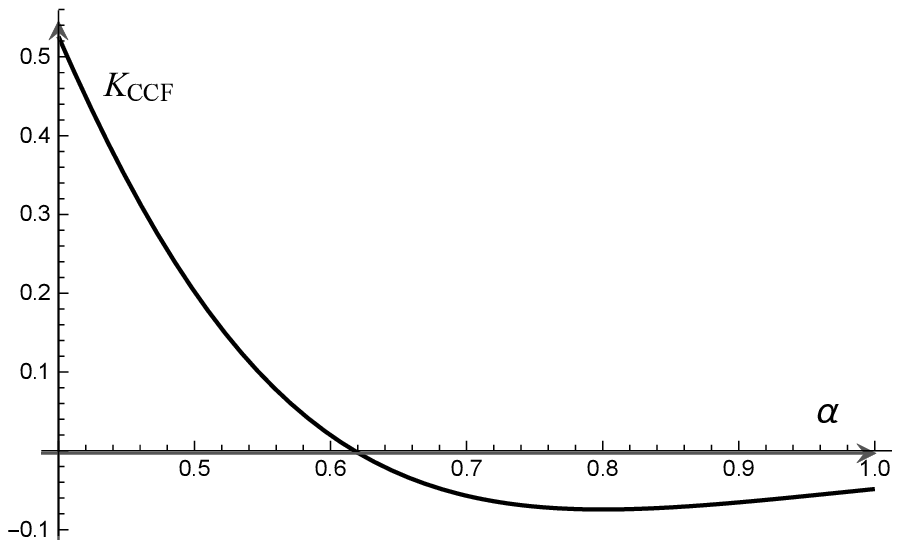}}
\caption{Spectral distribution associated with fractional C-C function,
	Eq.(\ref{artigo61}).}
\label{artigo33}
\end{figure}

\noindent If we choose $\alpha =1$ in the fractional C-C function given by
Eq.(\ref{artigo40}), we recover the fractional Debye function given by
Eq.(\ref{artigo9}). We can conclude that the fractional Debye function will be CM only if
$\gamma =1$, that is, in the integer case.
\begin{figure}[H]
\center
\subfigure[ref1][$\alpha=0.2$]{\includegraphics[width=7.5cm]{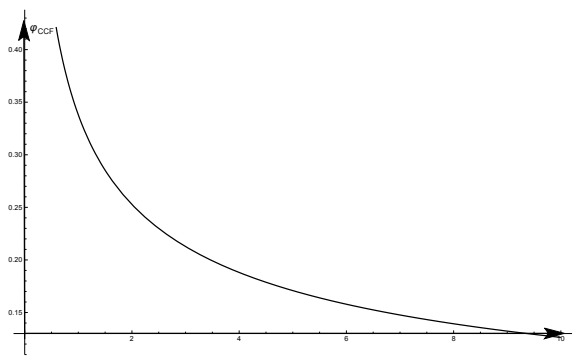}}
\qquad
\subfigure[ref2][$\alpha=0.3$]{\includegraphics[width=7.5cm]{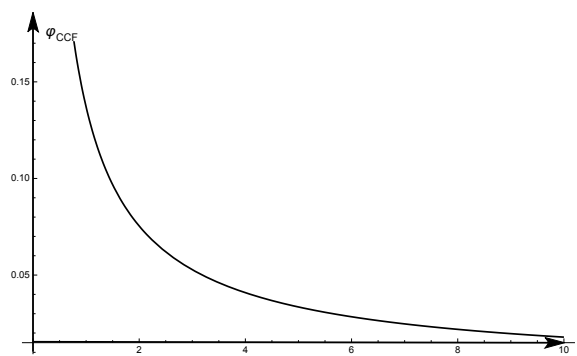}}
\qquad
\subfigure[ref1][$\alpha=0.7$]{\includegraphics[width=7.5cm]{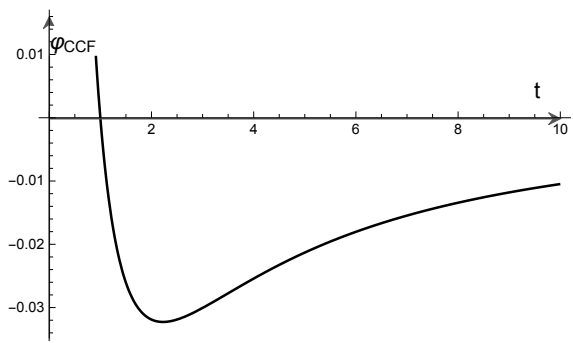}}
\qquad
\subfigure[ref4][$\alpha=0.9$]{\includegraphics[width=7.5cm]{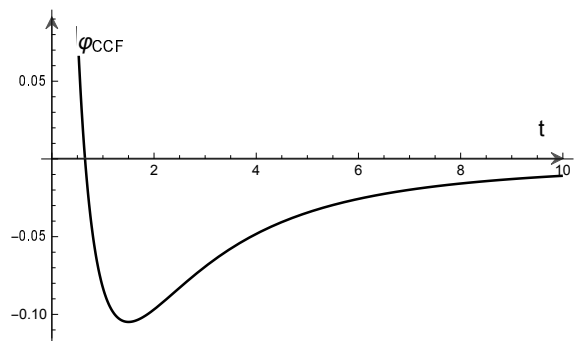}}
\caption{Function $\varphi_{CCF}(t)=
	t^{\gamma-1}E_{\alpha,\gamma}(-t^\alpha)$, for $\gamma=0.4$.}
\label{artigo7}
\end{figure}
\begin{figure}[H]
\center
\subfigure[ref1][$\alpha=0.2$]{\includegraphics[width=7.5cm]{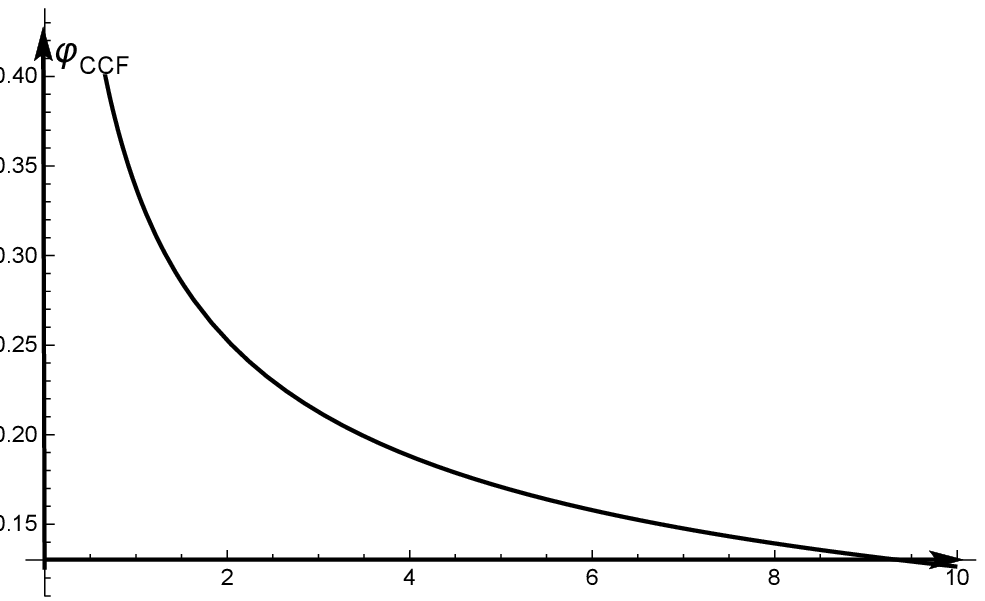}}
\qquad
\subfigure[ref2][$\alpha=0.6$]{\includegraphics[width=7.5cm]{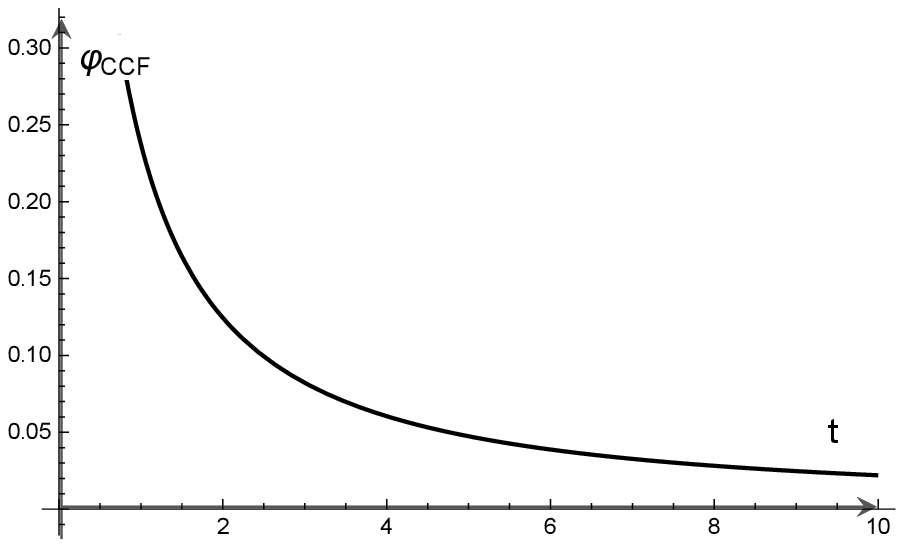}}
\qquad
\subfigure[ref1][$\alpha=0.8$]{\includegraphics[width=7.5cm]{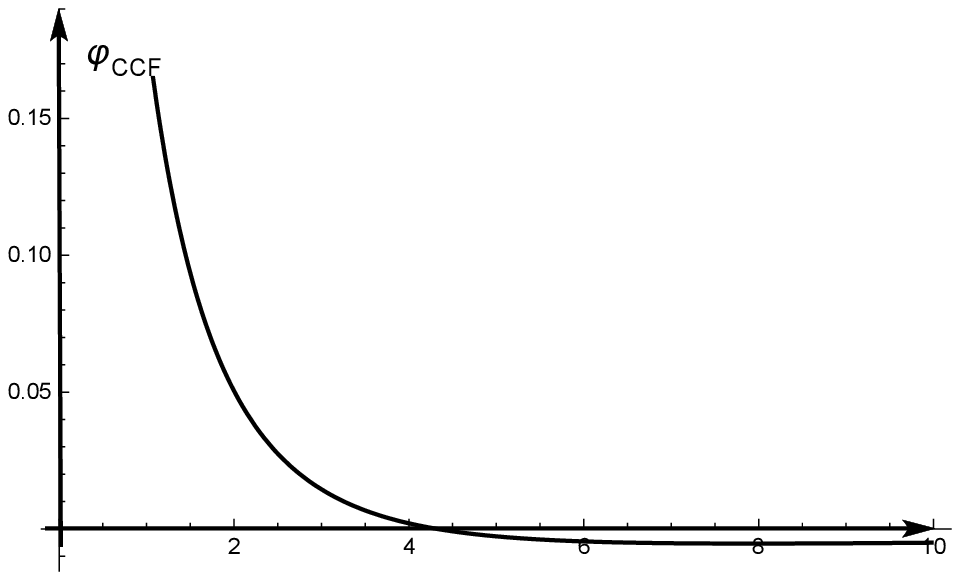}}
\qquad
\subfigure[ref4][$\alpha=0.9$]{\includegraphics[width=7.5cm]{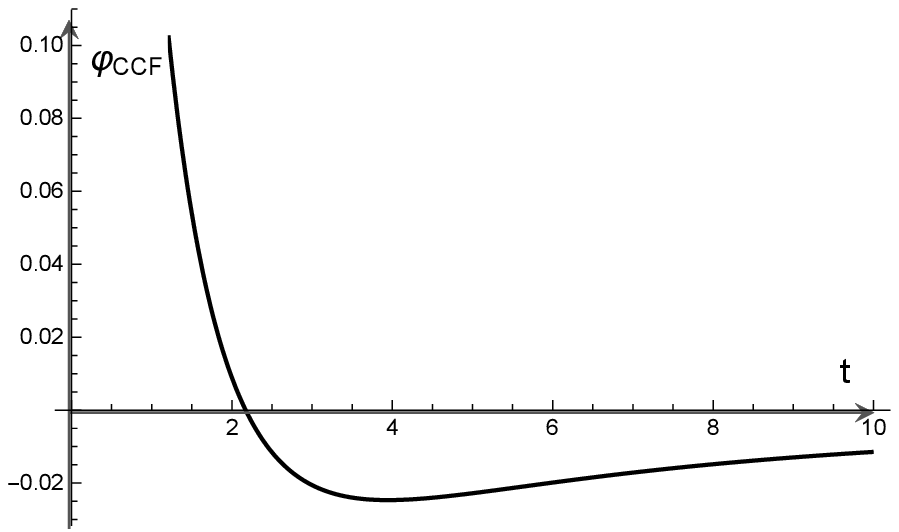}}
\caption{Function $\varphi_{CCF}(t)=
	t^{\gamma-1}E_{\alpha,\gamma}(-t^\alpha)$, for $\gamma=0.7$.}
\label{artigo8}
\end{figure}

\item Now, starting with the equation that generalizes the C-D model, that
	is, 
\begin{equation}
D^{\gamma}_{t}\varphi_{CDF}(t)=-
	D^{\gamma}_{t}\left\{M_{CD}\ast\varphi_{CDF}(t)\right\}=-
	D^{\gamma}_{t}\left\{\left[
	e^{-\frac{t}{\sigma}}\sigma^{-\beta}t^{\beta-1}E_{\beta,\beta}\left[-
	\left(\frac{t}{\sigma}\right)^{\beta}\right]
	\right]\ast\varphi_{CDF}(t)\right\},
\label{cdf1}
\end{equation}
we find that its soluction, which we call fractional C-D function, has the
following form: 
\begin{equation}
\varphi_{CDF}(t)=
	\frac{t^{\gamma-1}}{\Gamma(\gamma)}-t^{\beta+\gamma-1}
	E_{1,\beta+\gamma}^{\beta}(-t), \,\,\,\,\,\mbox{for}\,\,t\geq0.
\label{artigo28}
\end{equation}

As we can see, this function is formed by the difference between a power
function and a product of a power function by a Mittag-Leffler function
with three parameters. 

This function is not CM for $\gamma <1$ because we would then have $\alpha
= 1 > \gamma$ and as we shall show in the sequence, graphical evidence
suggests that when $\alpha$ is greater than $\gamma$, the function is not
CM. 


Figure \ref{figura9} shows the graphics of several particular cases of this
function. As we can see, the function assumes negative values very fast.

\begin{figure}[H]
\center
\subfigure[ref1][$\beta=0.4$ and $\gamma=0.2$]{\includegraphics[width=7.5cm]{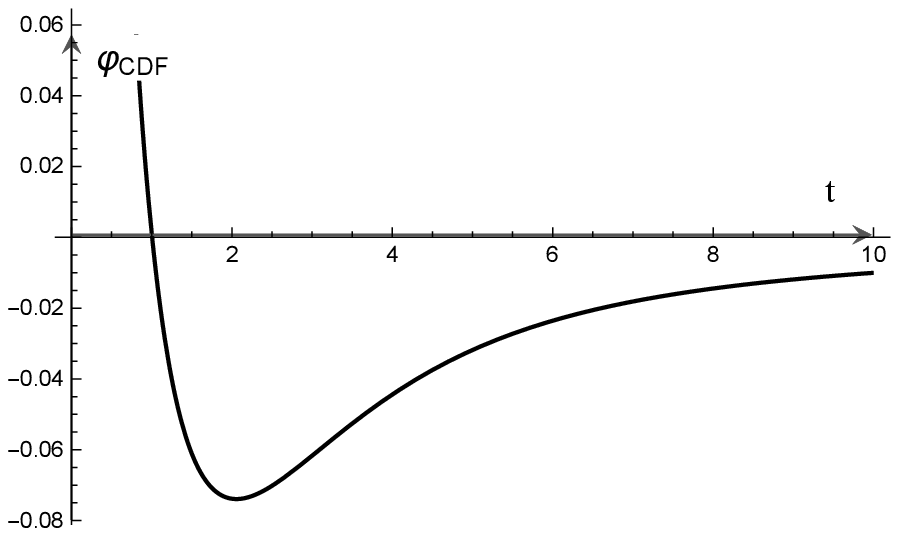}}
\qquad
\subfigure[ref1][$\beta=0.2$ and $\gamma=0.4$]{\includegraphics[width=7.5cm]{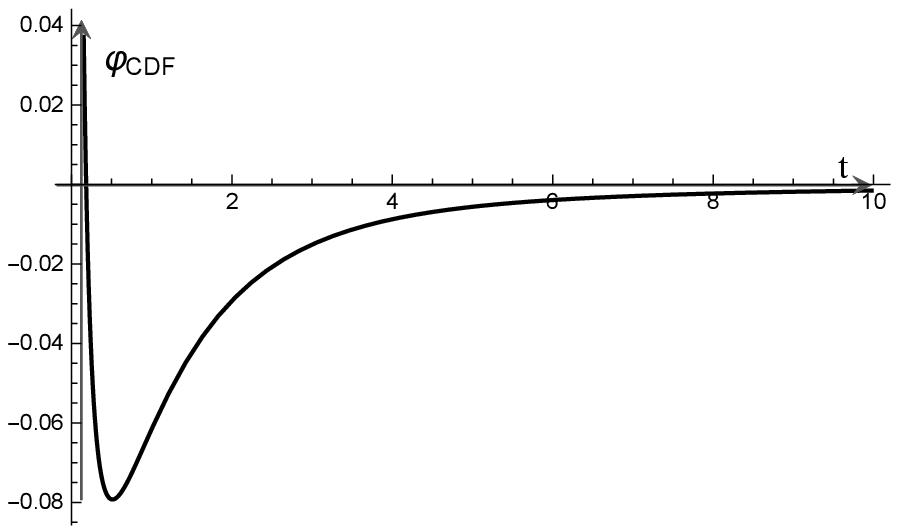}}
\qquad
\subfigure[ref1][$\beta=0.6$ and $\gamma=0.8$]{\includegraphics[width=7.5cm]{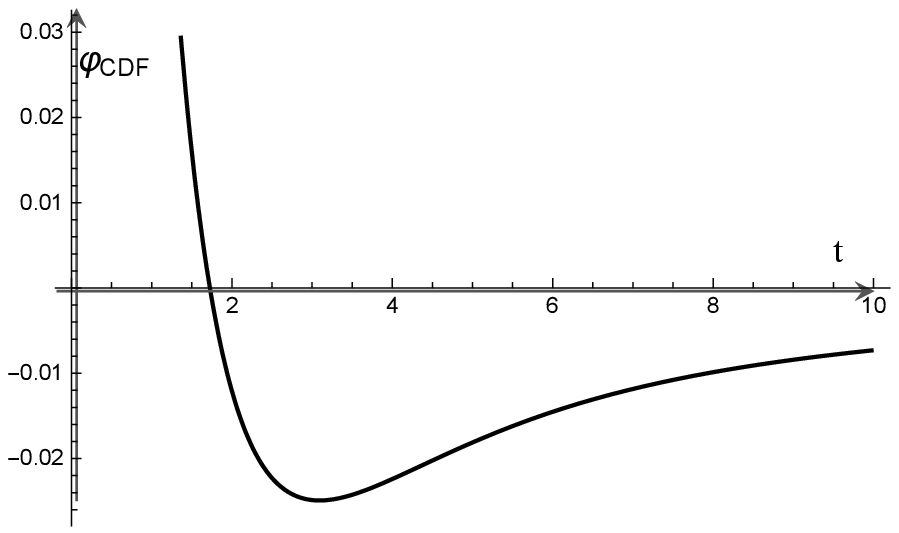}}
\qquad
\subfigure[ref1][$\beta=0.8$ and $\gamma=0.6$]{\includegraphics[width=7.5cm]{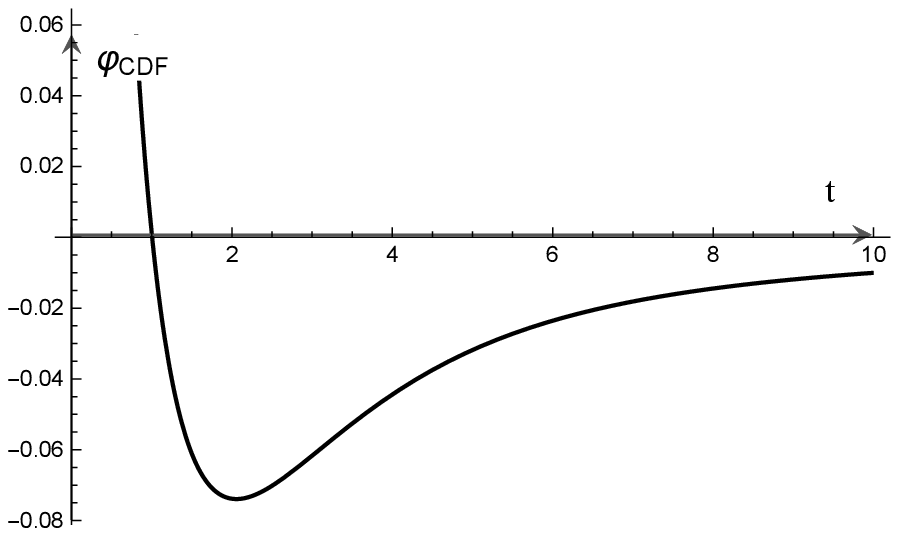}}
\qquad
\subfigure[ref1][$\beta=0.8$ and $\gamma=0.3$]{\includegraphics[width=7.5cm]{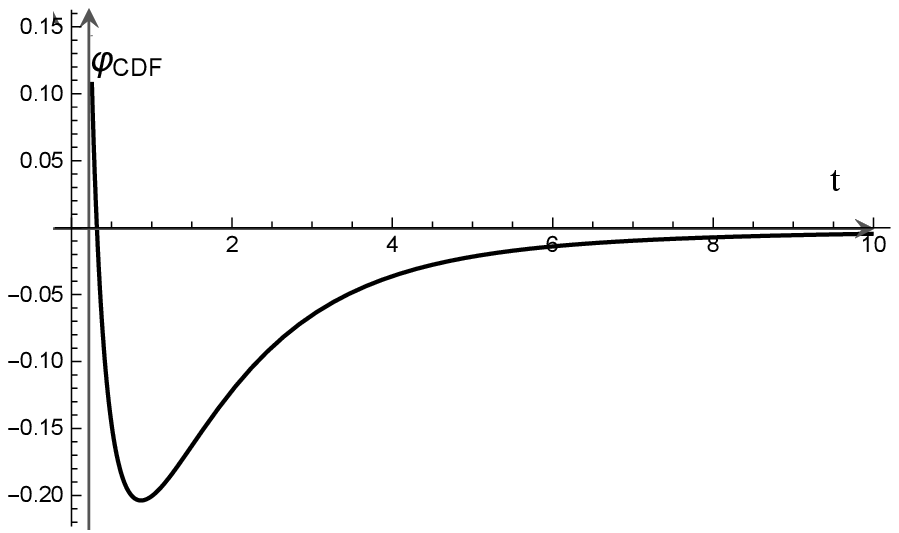}}
\qquad
\subfigure[ref1][$\beta=0.3$ and $\gamma=0.8$]{\includegraphics[width=7.5cm]{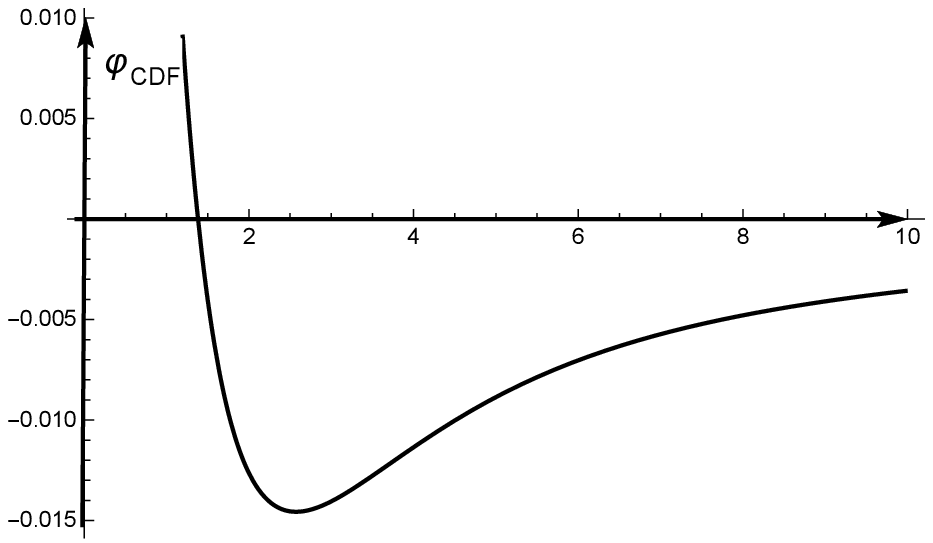}}
\caption{Function $\varphi_{CDF}(t)=
	\frac{t^{\gamma-1}}{\Gamma(\gamma)}
	-t^{\beta+\gamma-1} E_{1,\beta+\gamma}^{\beta}(-t)$.}
\label{figura9}
\end{figure}

\item Finally, let us consider the fractional kinetic relaxation equation 

\begin{equation}
D^{\gamma}_{t}\varphi_{HNF}(t)=-
	D^{\gamma}_{t}\left\{\left[\sum_{k=0}^{\infty}\sigma^{-\alpha
	k(\beta+1)}t^{\alpha\beta(k+1)-1}
	E_{\alpha,\alpha\beta(k+1)}^{\beta(k+1)}
	\left[-\left(\frac{t}{\sigma}\right)^{\alpha}\right]\right]
	\ast\varphi_{HNF}(t)\right\},
\label{hnf1}
\end{equation}
whose solution is the fractional H-N function given by 
\begin{equation}
\varphi_{HNF}(t)= \frac{t^{\gamma-1}}{\Gamma(\gamma)}
	-t^{\alpha\beta+\gamma-1}E_{\alpha,\alpha\beta+\gamma}^{\beta}
	(-t^\alpha),\,\,\,\,\,\mbox{for}\,\,t\geq0.
\label{artigo25}
\end{equation}

Let us now use Titchmarsh's formula to calculate the spectral distribution
of the fractional H-N function.  In order to do that, we need to find the
LT of the function given by Eq.(\ref{artigo25}). Writing
$\tilde{\varphi}_{HNF}(s) =\mathscr{L} [ \varphi_{HNF}(t)](s)$ we have 
\begin{equation}
\tilde{\varphi}_{HNF}(s)=\frac{1}{s^{\gamma}}-
	\frac{1}{s^{\gamma}(s^{\alpha}+1)^{\beta}}
\end{equation}
Hence, its distribution function is given by 
\begin{equation}
K_{HNF}=\frac{1}{\pi r^{\gamma}} \left[\sin(\pi \gamma)-
	\frac{\sin(\pi \gamma +\beta \theta)}{(r^{2\alpha}+2r^{\alpha}\cos(\alpha \pi)+1)^{\frac{\beta}{2}}}\right],
\label{artigo27}
\end{equation}
where  
\begin{equation}
\theta= \frac{\pi}{2}-\arctan\left[\frac{1+r^{\alpha}\cos(\alpha \pi)}{r^{\alpha}\sin(\alpha \pi)}\right].
\end{equation}

If we consider a very high value for $r$ in Eq.(\ref{artigo27}), the term
determining the function's sign will be $\sin (\pi \gamma)$; it is
positive, as $0<\gamma\leq1$. Thus, in order to investigate the variations
of sign of this distribution as a function of its parameters we have drawn
the curves shown in Figures \ref{artigo35} and \ref{artigo30} for a small
value of $r$.  We can see that this distribution function, though very
close to zero, becomes negative when parameter $\alpha$ is greater than
$\gamma$, the parameter of the fractional derivative, independently of the
value of parameter $\beta$. This also explains why the fractional C-D
function given by Eq.(\ref{artigo28}) is not CM when $\alpha=1$ and $0 <
\gamma <1$. We can thus conclude that the fractional C-D function is CM
only in the integer case $\gamma=1$. In the case $\alpha = \gamma =1$, as
we have seen, this function is CM for $0 < \beta \leq 1$.

\begin{figure}[H]
\center
\includegraphics[width=9cm]{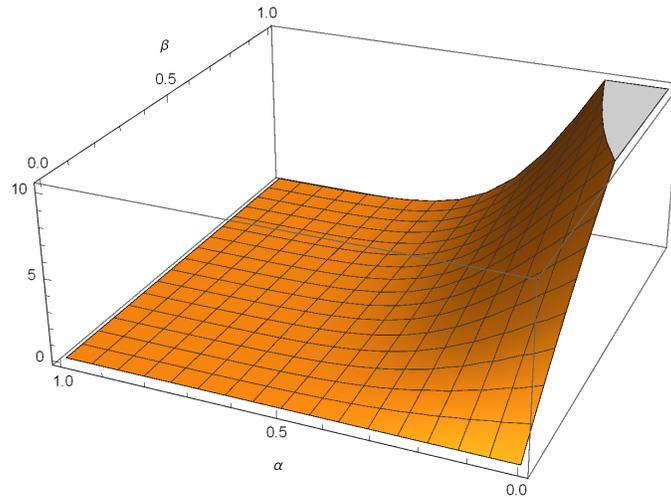}
\caption{Distribution function for fractional H-N function for 
	$\gamma=0.7$, according to Eq.(\ref{artigo27}).}
\label{artigo35}
\end{figure}
\begin{figure}[H]
\center
\subfigure[ref1][$\gamma=0.7$, $\beta=0.2$.]{\includegraphics[width=7.5cm]{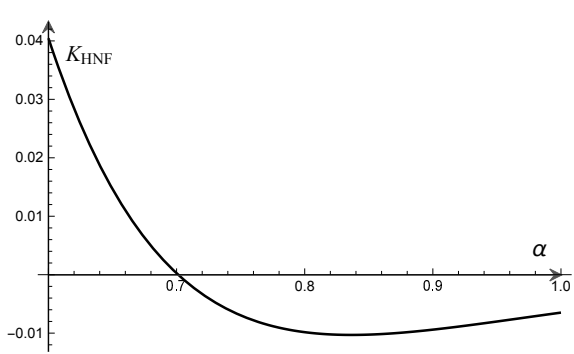}}
\qquad
\subfigure[ref2][$\gamma=0.3$, $\beta=0.8$.]{\includegraphics[width=7.5cm]{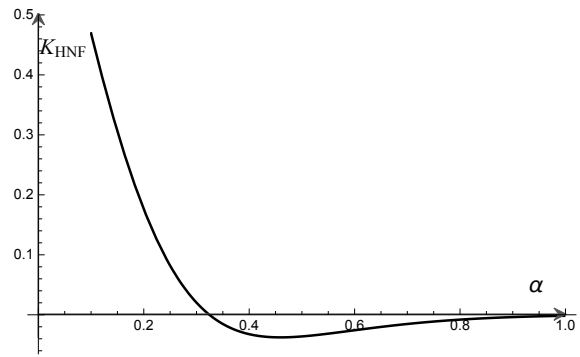}}
\qquad
\subfigure[ref3][$\alpha=0.4$,  $\beta=0.8$.]{\includegraphics[width=7.5cm]{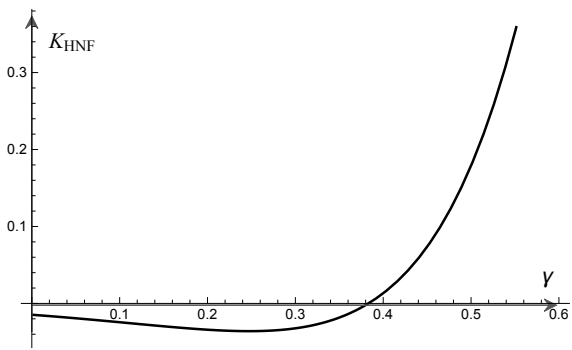}}
\qquad
\subfigure[ref4][$\alpha=0.8$, $\beta=0.5$.]{\includegraphics[width=7.5cm]{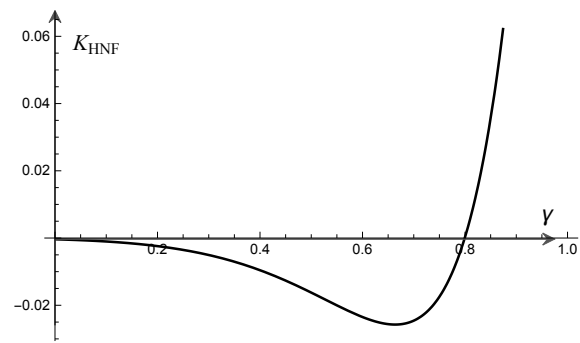}}
\caption{Spectral distribution function for the fractional H-N function,
	according to Eq.(\ref{artigo27}).}
\label{artigo30}
\end{figure}
\end{itemize}

\section{Conclusions}

The algebraic expressions and the graphic representations of relaxation
functions and fractional relaxation functions have shown that their complete
monotonicity depends on the parameters appearing in their definitions. We
thus analysed the variation of these parameters in order to determine the
conditions that should be imposed on them in order to ensure their complete
monotonicity. 

In the case of Debye's function, which is independent of those parameters,
its complete monotonicity follows directly from its definition. On the
other hand, in the case of relaxation functions associated with models C-D,
H-N and fractional D and C-C, the study of their complete monotonicity
follows from the trigonometric inequality arising from Titchmarsh inversion
formula for the spectral distribution function. The function appearing in the C-C model
is a particular case of a Mittag-Leffler function with three parameters;
its complete monotonicity was demonstrated in the first part of this work,
using, among other results, Theorem (\ref{teoremaimportante}). In the case
of the relaxation function associated with the fractional H-N model, the
associated spectral distribution function has values very close to zero.
Nevertheless, the study of its complete monotonicity was based on the
graphical analysis of this distribution function. The conclusion about the
complete monotonicity of the fractional C-D function is immediate as it is
a particular case of fractional H-N function.

\bibliography{ref}
\bibliographystyle{plain}

\end{document}